\documentclass[onecolumn,aps,amsmath,amssymb,floatfix,prl,superscriptaddress,showkeys]{revtex4}

\usepackage{bbm}
\usepackage[dvips]{graphicx}
\usepackage{xcolor} 

\usepackage{amsmath,amssymb,latexsym}

\renewcommand{\ol}[1]{\overline{#1}}

\newcommand{\F}{\mathbf{F}}
\renewcommand{\L}{\mathcal{L}}
\newcommand{\M}{\mathbf{M}}

\newcommand{\Q}{\mathbf{Q}}
\newcommand{\R}{\mathcal{R}}
\renewcommand{\S}{\mathcal{S}}

\newcommand{\e}{\mathbf{e}}
\newcommand{\f}{\mathbf{f}}
\newcommand{\g}{\mathbf{g}}

\newcommand{\I}{\mathcal{I}}

\newcommand{\n}{\mathbf{n}}
\newcommand{\q}{\mathbf{q}}
\renewcommand{\r}{\mathbf{r}}
\renewcommand{\u}{\mathbf{u}}
\renewcommand{\v}{\mathbf{v}}
\newcommand{\w}{\mathbf{w}}
\newcommand{\x}{\mathbf{x}}
\newcommand{\X}{\mathbf{X}}

\newcommand{\act}{\mathrm{act}}
\newcommand{\ext}{\mathrm{ext}}
\newcommand{\micron}{\mu\mathrm{m}}

\newcommand{\stress}{\boldsymbol{\sigma}}
\newcommand{\strain}{\boldsymbol{\Delta}}
\newcommand{\zerovec}{\boldsymbol{0}}
\newcommand{\onevec}{\boldsymbol{1}}
\newcommand{\comma}{\quad,\quad}
\newcommand{\Omegavec}{\boldsymbol{\Omega}}
\newcommand{\nablavec}{\boldsymbol{\nabla}}

\definecolor{brown}{rgb}{0.5,0,0}

\newcommand{\diff}{d} 
\begin{document}

\title{Lagrangian Mechanics of Active Systems}

\author{Anton Solovev}
\affiliation{TU Dresden, Dresden, Germany}
\author{Benjamin M. Friedrich}
\email{benjamin.m.friedrich@tu-dresden.de}
\affiliation{TU Dresden, Dresden, Germany}

\date{\today}


\begin{abstract} 
We present a multi-scale modeling and simulation framework for low-Reynolds number hydrodynamics of 
shape-changing immersed objects, e.g., biological microswimmers and active surfaces.
The key idea is to consider principal shape changes as generalized coordinates, 
and define conjugate generalized hydrodynamic friction forces.
Conveniently, the corresponding generalized friction coefficients can be pre-computed 
and subsequently re-used to solve dynamic equations of motion fast.
This framework extends Lagrangian mechanics of dissipative systems to active surfaces and active microswimmers, 
whose shape dynamics is driven by internal forces.
As an application case, we predict in-phase and anti-phase synchronization in pairs of cilia for an experimentally measured cilia beat pattern. 
\end{abstract}

\maketitle

\paragraph{Biological hydrodynamics.}
Biology provides ample examples of active shape-changes in fluid environments: 
bacteria like \textit{E. coli} rotate helical prokaryotic flagella to swim \cite{Berg1973}, 
other bacteria like \textit{Spiroplasma} propagates twist waves along their flexible body \cite{Wada2007},
sperm cells and motile algae posses slender cell appendages termed cilia (or eukaryotic flagella), 
whose regular bending waves propel these cells in a fluid \cite{Gray1928,Gray1955a}.
On epithelial surfaces, collections of beating cilia
transport biological fluids such as mucus in airways, cerebrospinal fluid in brain ventricles, and oviduct fluid in the Fallopian tubes
\cite{Sanderson1981,Faubel2016}.
In addition to their important role in self-propulsion and fluid transport,
these model systems enable us to learn about internal force generation mechanisms in these cells, 
such as the collective dynamics of molecular motors inside cilia \cite{Brokaw1972,Lindemann1994,Riedel2007,Klindt2016}. 
On larger scales, the interaction of many shape-changing units leads to the spontaneous formation of spatio-temporal patterns, 
e.g., in dense suspensions of microswimmers \cite{Riedel2005},
or collections of cilia exhibiting metachronal coordination \cite{Machemer1972}.
 
These examples represent a class of fluid-structure interaction problems, 
where shape-changing active structures exert forces on the surrounding fluid, 
while the surrounding passive fluid exerts hydrodynamic friction forces back on these active structures.
These hydrodynamic forces may affect the active shape dynamics;
examples include the torque-velocity relationship of rotating prokaryotic flagella \cite{Berg1993}, 
the load-response of beating cilia and eukaryotic flagella \cite{Brokaw1966,Klindt2016}, 
as well as minimal model swimmers \cite{Golestanian2008,Pickl2017,Friedrich2018}.
Closed feedback loops between passive fluids and active structures
can lead to emergent dynamics;
examples include spontaneous pattern formation in dense microswimmer suspensions \cite{Riedel2005,Wensink2012}, 
or (hydrodynamic) synchronization of beating cilia and flagella \cite{Machemer1972,Ruffer1998a,Goldstein2009,Woolley2009,Brumley2014,Pellicciotta2020}.

\paragraph{Common hydrodynamics methods at low Reynolds numbers.}
At the relevant length and time scales, viscous drag dominates inertia, 
corresponding to low Reynolds numbers \cite{Purcell1977,Lauga2009,Elgeti2015}.
In the limit of zero Reynolds numbers, 
the Navier-Stokes equation of hydrodynamics simplifies to the Stokes equation. 
Although, the Stokes equation is linear, hydrodynamic computations can still be costly, 
because hydrodynamic interactions are long-ranged \cite{Happel:hydro}.

In the past, different computational methods of different degrees of approximation have been used in the community, 
including
resistive force theory for slender filaments, which includes short-range, but not long-range hydrodynamic interactions
\cite{Gray1955b,Johnson1979,Friedrich2010}, 
the more refined method of slender-body theory, which considers a line distribution of hydrodynamic singularities (point forces) along a filament
\cite{Batchelor1970,Keller1976,Smith2009},
or multi-particle collision dynamics, 
which replaces the continuum description of the Stokes equation
by the stochastic dynamics of a large number of ``fluid particles'' \cite{Gompper2009,Elgeti2008,Winkler2016}. 
Despite its applicability for large-scale problems \cite{Westphal2014}, 
the stochastic nature of the MPCD algorithm introduces algorithm-specific fluctuations, 
which can be impractical if one wants to study the role of biological noise.
Lattice-Boltzmann methods similarly rely on fictitious ``fluid particles'', 
for which in each time step both a streaming and a collision steps is performed \cite{Chen1998}.
Finally, boundary element methods convert the problem of solving the Stokes equation in three-dimensional space
to a two-dimensional boundary integral problem of finding a surface distribution of forces on a moving boundary surface.
Boundary element methods are similar in spirit to slender-body methods, 
but less susceptible to issues of regularization, since a two-dimensional distribution of forces is used.
Modern algorithms use fast multi-pole methods that 
solve a tree of hierarchically coarse-grained sub-problems
instead of solving a single large linear system when computing the force distribution on a surface \cite{Liu2006,Liu2009,Giuliani2020}.

Irrespective of the hydrodynamic computation method used, 
it can be computationally costly to calculate a solution of the Stokes equation in every time step, 
while simulating the dynamics of a shape-changing microswimmer or an active surface.

\paragraph{Lagrangian mechanics.}
In this methods manuscript, we present a multi-scale simulation framework, 
where the Stokes equation has to be solved only in an initial step for a small set of principal shape modes of a shape-changing surface.
The resultant surface distributions of hydrodynamic friction forces 
define generalized hydrodynamic friction coefficients
by a projection method of Lagrangian mechanics
\cite{Goldstein:mechanics,Vilfan2009,Friedrich2012,Geyer2013,Polotzek2013,Klindt2015,Klindt2016,Klindt2017}.
These scalar friction coefficients are independent of the velocity of the moving surface.
Once tabulated, these friction coefficients provide a look-up table for subsequent fast simulations of shape dynamics and active motion.
Specifically, we view principal shape changes of an active surface as generalized coordinates, 
for which we compute conjugate generalized friction forces.
We obtain effective equations of motion for the generalized coordinates
from a force balance between these generalized friction forces and active driving forces.
These active driving forces coarse-grain the internal active processes that drive the active shape changes of the surface
(such as the collective dynamics of molecular motors).
Importantly, these \textit{a priori} unknown active driving forces can be calibrated for a reference case 
(e.g., using experimental data), and then used to extrapolate to other application cases of interest.
Thereby, our framework extends Lagrangian mechanics of dissipative systems to active surfaces and active microswimmers, 
whose shape dynamics is driven by active forces.

\section{Notation: Stokes equation and hydrodynamic dissipation}

Fluid dynamics at the scale of individual biological cells is characterized by low Reynolds numbers, 
i.e., viscous effects commonly dominate over inertia \cite{Purcell1977,Lauga2009,Elgeti2015}.
Correspondingly, fluid flow is described by the \textit{Stokes equation}, 
which reads for an incompressible Newtonian fluid in the absence of body forces in the bulk \cite{Happel:hydro}
\begin{equation}
\label{eq:stokes}
\zerovec = - \nablavec p + \mu\,\nablavec^2\u \quad,
\end{equation}
with incompressibility condition $\nablavec\cdot\u = \zerovec$.
Here, $\u(\x)$ denotes the flow velocity, $p(\x)$ the pressure field, and $\mu$ the dynamic viscosity of the fluid. 

The total stress tensor $\stress$ for an incompressible fluid depends on both 
the hydrostatic pressure $p$ and the symmetrized strain rate tensor $\strain$ \cite{Happel:hydro}
\begin{equation} 
\label{eq:stress}
\stress = -p\,\onevec + 2\mu\,\strain
\quad,\quad
\strain = \frac{1}{2} \left[ \nablavec\otimes\u + (\nablavec\otimes\u)^T \right]
\quad.
\end{equation} 
Thus, the Stokes equation, Eq.~(\ref{eq:stokes}) could be equivalently written as 
$\zerovec = \nablavec\cdot\stress$ in the bulk of the fluid.
Special conditions apply at boundaries.

\paragraph{No-slip boundary condition for an active surface.}

We consider a surface $\S$ immersed in the fluid that changes its shape as a function of time.
For example, $\S$ may represent the outer surface of a shape-changing microswimmer, 
or even the combined surface for a collection of microswimmers.
We introduce the surface velocity $\v(\x,t)$ for each point $\x\in\S$ at time $t$.

We impose a \textit{no-slip boundary condition} at this surface,
i.e., require that the local velocity $\u(\x)$ of fluid flow
matches the local velocity $\v(\x)$ of the surface for each surface point
\begin{equation}
\u(\x) = \v(\x) \text{ for all } \x\in\S
\quad.
\end{equation}

\paragraph{Hydrodynamic friction forces.}

A shape change of the surface $\S$ induces a flow field $\u(\x)$ with corresponding stress tensor field $\stress(\x)$.
The stress $\sigma(\x)$ determines the surface density of forces $\f(\x)$ exerted \textit{by the surface on the fluid} 
(with units of a stress $\mathrm{N/m^2}$, also called contact force, or traction force density)
\begin{equation}
\f(\x) = -\stress\cdot\n
\text{ for all }
\x\in\S 
\quad,
\end{equation}
where $\n$ is the surface normal pointing into the fluid.
Correspondingly, $-f(\x)$ is the surface density of hydrodynamic friction forces exerted \textit{by the fluid on the surface}.
The total force exerted by the surface on the fluid is simply the surface integral of $\f(\x)$
\begin{equation}
\F = \int_{\S} \! d^2\x\, \f(\x) \quad.
\end{equation}
Analogously, the total torque (with respect to a reference point $\x_0$)
exerted by the surface on the fluid is given by
\begin{equation}
\M = \int_{\S} \! d^2\x\, (\x - \x_0) \times \f(\x) \quad.
\end{equation}

\paragraph{Superposition principle.}
The linearity of the Stokes equation of low-Reynolds number flow, Eq.~(\ref{eq:stokes}), 
implies a \textit{superposition principle} for hydrodynamic friction forces, 
which will be pivotal for the modeling ansatz presented here.
Specifically, we consider a boundary condition with rate of displacement $\v$ that is given as a linear combination of
velocity distributions $\v_1$ and $\v_2$ as 
\begin{equation}
\label{eq:superposition}
\v=\alpha_1\, \v_1 + \alpha_2\, \v_2 \quad,
\end{equation}
with real coefficients $\alpha_1, \alpha_2\in \mathbbm{R}$.
Then, 
the resultant flow field $\u$ is given by $\u=\alpha_1\u_1 + \alpha_2\u_2$, 
while the surface density of hydrodynamic friction forces $\f$ is $\f=\alpha_1\f_1+\alpha_2\f_2$, 
where $\u_i$ and $\f_i$ denote the flow field and the surface density of hydrodynamic friction forces 
corresponding to boundary condition $\v_i$, respectively, for $i=1,2$.

\paragraph{Hydrodynamic dissipation.}

We introduce the rate of work $\R^{(h)}$ exerted by the surface on the fluid
\begin{equation}
\label{eq:Rh}
\R^{(h)} = \int_{\S} \! d^2\x\, \v(\x)\cdot\f(\x) \quad.
\end{equation}
For incompressible Newtonian fluids at zero Reynolds number, 
$\R^{(h)}$ equals the instantaneous rate of hydrodynamic energy dissipation within the fluid \cite{Happel:hydro}.
Indeed, 
let us consider
the local dissipation rate, which is given by 
$\Phi = 2\mu\, {\strain}:{\strain}$, 
where $\strain:\strain=\sum_{i,j} \Delta_{ij}\Delta_{ij}$ denotes tensor contraction.
The dissipation rate can be rewritten as 
$\Phi = \nablavec\cdot(\u\cdot\stress)$ using Eqs.~(\ref{eq:stokes}), (\ref{eq:stress}) 
and the incompressibility condition $\nablavec\cdot\u=\zerovec$.
Gauss divergence theorem now gives \cite{Happel:hydro}
(using $\u(\x)=\v(\x)$ for $\x\in\S$)
\begin{equation}
\label{eq:Rh_Gauss}
\underbrace{ 
\int_{\S} \! d^2\x\, \v(\x)\cdot\f(\x)
}_\text{power exerted by surface} 
= 
\underbrace{
\int_V \! d^3\x\, \Phi(\x)
}_\text{hydrodynamic dissipation in bulk}
\quad.
\end{equation}
Here, $V$ denotes the three-dimensional fluid domain with boundary surface $\S$.
At finite Reynolds numbers, $\R^{(h)}$ still equals the rate of work exerted by the surface on the fluid, 
yet this injected energy would be dissipated as heat with a delay, 
such that Eq.~(\ref{eq:Rh_Gauss}) would only hold for time-averages.

\section{Lagrangian mechanics: Generalized coordinates}

We consider a shape-changing surface $\S(t)$.
While a description of all possible shape changes of $\S$ would require an infinite number of degrees of freedom, 
in important application cases,
we can restrict ourselves to a constrained set of shape changes characterized by a small number of shape coefficients, 
or generalized coordinates, $q_1,\ldots,q_n$.

\begin{figure}
\includegraphics[width=0.7\linewidth]{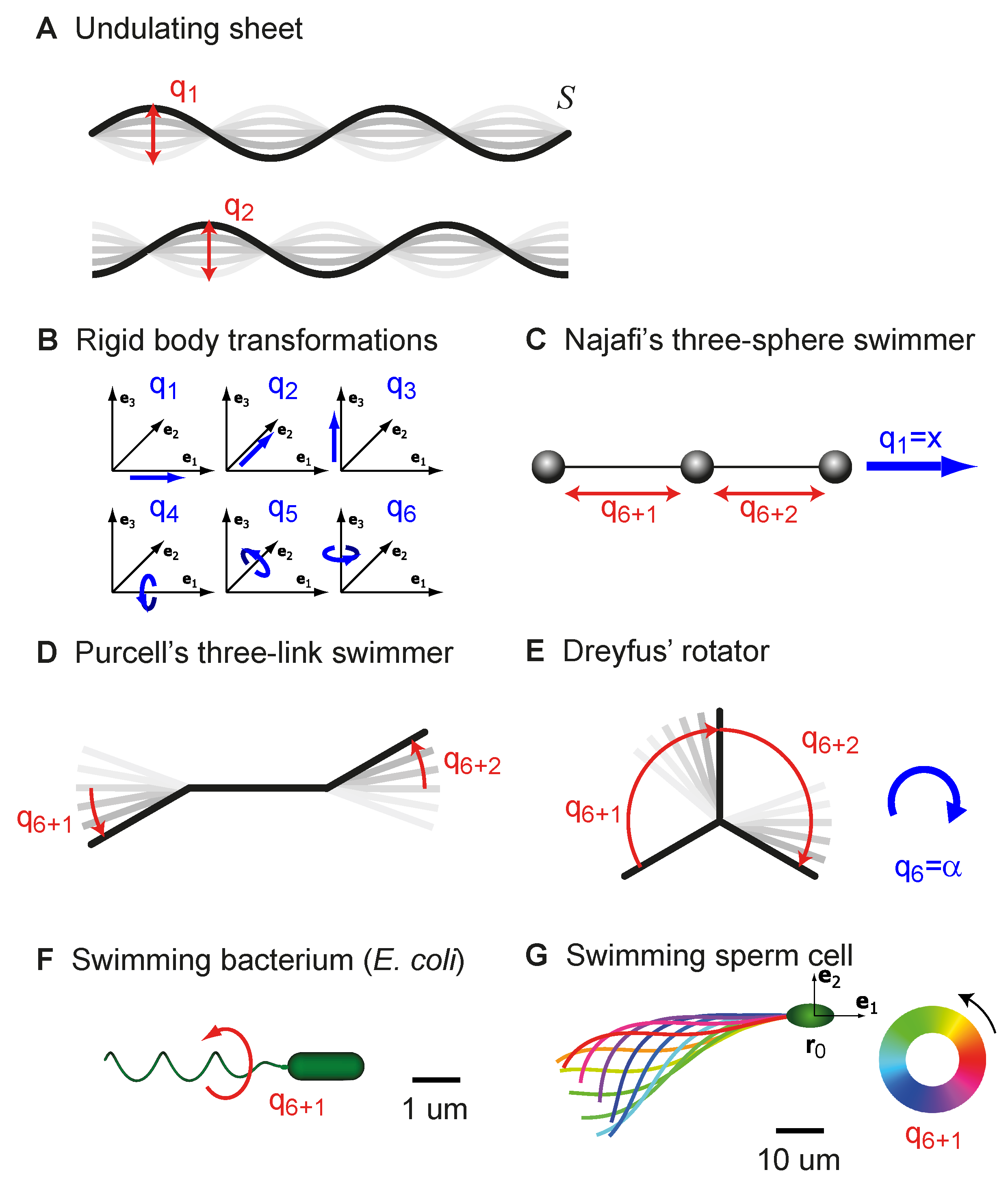} 
\caption{
\textbf{Generalized coordinates: Examples.}
(A) 
Undulating sheet with two wave modes.
The amplitudes $q_1$, $q_2$ of the wave modes represent generalized coordinates of the shape-changing surface $\S$.
(B) 
Rigid body motion of a microswimmer in three-dimensional space 
is characterized by three translational and three rotational degrees of freedom, 
corresponding to six generalized coordinates: 
$q_i$ for translations parallel to the $\e_i$-axis, and 
$q_{i+3}$ for rotations around the $\e_i$-axis, $i=1,2,3$, respectively. 
(C) 
Najafi's three-sphere swimmer consists of three collinear spherical beads with time-varying bead distances \cite{Najafi2004},
corresponding to two internal degrees of freedom, $q_{6+1}$ and $q_{6+2}$, 
in addition to the generalized coordinates of rigid body motion.
(D) 
Purcell's three-link swimmer consists of three connected segments \cite{Becker2003}, 
whose relative angles $q_{6+1}$ and $q_{6+2}$ can be treated as two generalized coordinates.
(E)
Similarly, Dreyfus' rotator consists of three segments connected at a single joint;
the relative angles $q_{6+1}$ and $q_{6+2}$ again define generalized coordinates.
This shape-changing microswimmer exhibits pronounced rotation in the plane in addition to translational motion, hence its name.
(F)
Simplified geometry of the bacterium \textit{E. coli} with a single prokaryotic flagellum.
A rotary motor inside the cell wall can spin the helical flagellum around its central axis;
this internal rotational degree of freedom defines a single generalized coordinate $q_{6+1}$ with periodicity of $2\pi$. 
(G)
Prototypical flagellar beat pattern of a sperm cell, parametrized by a $2\pi$-periodic phase variable, 
which defines a generalized coordinate $q_{6+1}$. 
For the amplitude of regular flagellar bending waves and mean flagellar curvature, 
we used parameters from \cite{Friedrich2010}.
}
\label{figure1}
\end{figure}

Examples for minimal model swimmers include 
undulating sheets with a finite set of admissible wavelengths \cite{Taylor1951},
bead distances as in Najafi's three-sphere swimmer \cite{Najafi2004}, or
lever arm angles in Purcell's the three-link swimmer \cite{Becker2003} and Dreyfus' rotator \cite{Dreyfus2005}, 
see Fig.~\ref{figure1}.
An example for a biological microswimmer would be the 
rotation angle $\varphi$ of an idealized rigid helical prokaryotic flagellum.
Similarly, the regular traveling bending waves of cilia and eukaryotic flagella 
can be described by an oscillator phase $\varphi$ that characterizes the current position in a periodic shape cycle
\cite{Geyer2013,Ma2014,Wan2014,Werner2014}.
Elastic degrees of freedom arising from waveform compliance can be incorporated in such a framework as additional amplitude degrees of freedom
\cite{Klindt2016,Klindt2017}.

We introduce the \textit{state vector}, $\q=(q_1,\ldots,q_n)$.
The shape dynamics of the active surface $\S(t) = \S[\q(t)]$ is thus entirely described by the dynamics of $\q(t)$. 
In particular, the local rate of surface displacement depends linearly on the generalized velocities $\dot{q}_i$ as 
\begin{equation}
\label{eq:v_superposition}
\v(\x) = \w_1(\x;\q)\, \dot{q}_1 + \w_2(\x;\q)\, \dot{q}_2 + \ldots + \w_n(\x;\q)\, \dot{q}_n \quad,
\end{equation}
where the normalized velocity fields 
$\w_i(\x) = \partial \x / \partial q_i$ 
depend on $\q(t)$ but not on $\dot{\q}(t)$.
In fact, 
Eq.~(\ref{eq:v_superposition})
simply generalizes Eq.~(\ref{eq:superposition}) 
to the case of generalized coefficients $\alpha_i = \dot{q}_i$ with units of a generalized velocity.
Correspondingly, the surface distribution of hydrodynamic friction forces $\f(\x)$ 
is given as a linear combination
\begin{equation}
\label{eq:f_superposition}
\f(\x) = \g_1(\x;\q)\, \dot{q}_1 + \ldots + \g_n(\x;\q)\,\dot{q}_n \quad,
\end{equation} 
where 
the normalized force densities
$\g_i(\x;\q) = \f_i(\x)/\alpha_i $ 
correspond to the surface density of hydrodynamic friction forces
$\f_i(\x)$
induced by the velocity field 
$\v_i(\x) = \alpha_i\, \w_i(\x;\q)$.
An example of a surface velocity field with corresponding surface density of hydrodynamic friction forces 
is shown in Fig.~\ref{figure2}.

The formalism allows to include also rigid body transformation such as translations and rotations of the surface $\S$
in the set of generalized coordinates.
Thereby, the self-propulsion of shape-changing microswimmers can be described using the same formalism, 
see the section of rigid body transformations below.

\section{Generalized hydrodynamic friction forces}

We introduce generalized hydrodynamic friction forces $P_i$ conjugate to the generalized coordinates $q_i$, 
following the convention of Lagrangian dynamics of dissipative systems \cite{Goldstein:mechanics},
see also \cite{Vilfan2009,Polotzek2013}
\begin{equation}
\label{eq:P_i}
P_i = \int_{\S} \!d^2\x\, \w_i(\x) \cdot \f(\x)
\comma
i=1,\ldots,n
\quad.
\end{equation}
The superposition principle for the shape changes $\w_i(\x)$, 
Eq.~(\ref{eq:v_superposition}), 
allows us to rewrite the total hydrodynamic dissipation rate $\R^{(h)}$ 
as a sum of products of generalized velocities times their conjugate generalized friction force
\begin{equation}
\R^{(h)} = \sum_i P_i\, \dot{q}_i\quad.
\end{equation}
Note that the different generalized coordinates $q_i$ may have different physical units, 
in which case also all derived quantities will have different units;
nonetheless, all vector and matrix operations of the formalism are consistent unit-wise.

In the special case, where some of the $q_i$ denote a rigid body transformation of an immersed microswimmer, 
i.e., a rigid body translation or rotation, the conjugate generalized force simply corresponds to the respective components 
of the total force $\F$ or total torque $\M$ exerted by the swimmer on the fluid, respectively, 
see the section on rigid body motion below. 

\paragraph{Generalized hydrodynamic friction coefficients.}

\begin{figure}
\includegraphics[width=0.7\linewidth]{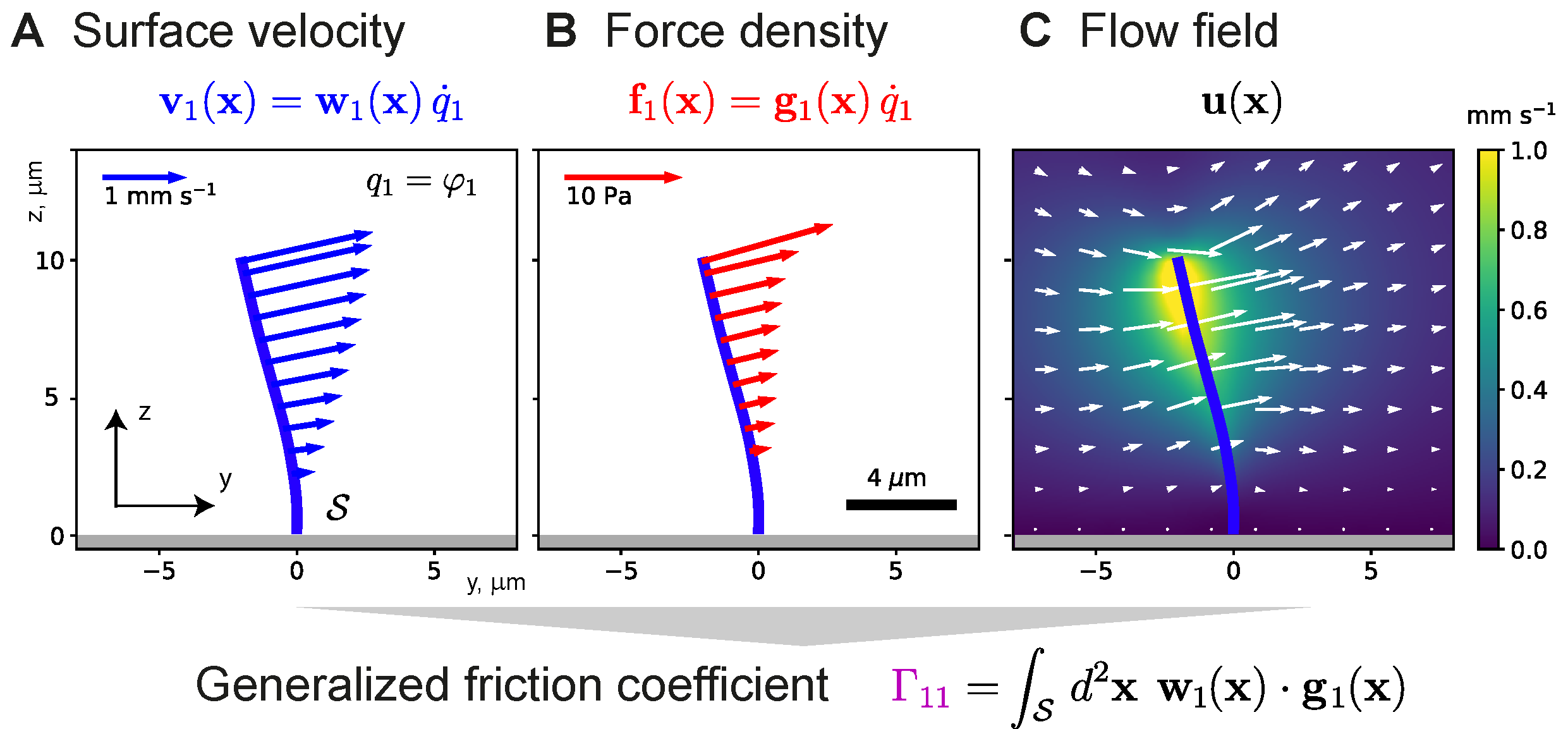}
\caption{
\textbf{Hydrodynamic friction forces: Example of cilium during power stroke.}
(A)
Surface velocity $\v_1(\x)$ on a shape-changing surface $\S$, 
here given by a slender cilium (blue) attached to a no-slip boundary surface (gray);
the cilium progresses with phase velocity $q_1=\dot{\varphi}_1$ along its periodic beat cycle. 
For visualization, the three-dimensional shape of the cilium, as well as $\v_1(x)$ was projected on the $yz$-plane
(see Fig.~\ref{figure3}A for a three-dimensional representation).
(B) 
Corresponding surface distribution of hydrodynamic friction forces $\f_1(\x)$ 
exerted by the active, shape-changing surface on the surrounding viscous fluid.
The force distribution is obtained by solving the Stokes equation, Eq.~(\ref{eq:stokes}), 
see also \textit{Multi-scale modeling} section for details.
From the velocity and force distributions, $\v_1(\x)$ and $\f_1(\x)$, 
we can compute a generalized hydrodynamic friction coefficient, 
$\Gamma_{11}$, 
which is proportional to the phase-dependent rate of energy dissipation in the surrounding fluid.
In the general case of $n$ generalized coordinates $q_1,\ldots,q_n$, we obtain a $n\times n$-matrix $\Gamma_{ij}$, 
see also Eq.~(\ref{eq:Gamma}).
(C)
Flow field induced by the active shape change of the cilium, here shown as two-dimensional section at $x=0$.  
The color represents the magnitude $|\u(\x)|$ of three-dimensional velocity vectors, 
whereas white arrows represent the projections of $\u$ on the $yz$-plane.
The flow field was computed as 
convolution 
of the fundamental solution of the Stokes equation with the force distribution $\f_1(\x)$.
Cilium phase corresponding to Fig.~\ref{figure3}A:
 $\varphi_1 = 1.4\,\pi$,
cilium beat frequency: $\omega_0 /(2\pi) = 32 \,\mathrm{Hz}$ \cite{Machemer1972},
dynamic viscosity of fluid: $\mu = 10^{-3}\, \textrm{Pa}\,\textrm{s}$ 
(corresponding to viscosity of water at $20^\circ\,\mathrm{C}$).
}
\label{figure2}
\end{figure}

Using the superposition principle of Stokes flow, 
we can conveniently express the generalized hydrodynamic friction forces 
as linear function of the generalized velocities $\dot{q}_i$ 
by introducing \textit{generalized hydrodynamic friction coefficients}
\begin{equation}
P_i = \sum_{j=1}^n \Gamma_{ij}\, \dot{q}_j \comma
i=1,\ldots,n \quad. 
\end{equation}
The generalized friction coefficients can be computed as 
scalar products between 
the (normalized) velocity profiles $\w_i(\x)$, and
the (normalized) force profiles $\g_j(\x)$, see also Fig.~\ref{figure2}
\begin{equation}
\label{eq:Gamma}
\Gamma_{ij} =
\int_{\S} \! d^2\x\, 
\w_i(\x) \cdot \g_j(\x) 
\comma
i,j=1,\ldots,n \quad. 
\end{equation}
Alternatively, we could express $\Gamma_{ij}$ 
in terms of partial derivatives with respect to the generalized velocities $\dot{q}_i$ as
$\Gamma_{ij} = \int_{\S} \! d^2\x\, (\partial\,\v(\x)/\partial\,\dot{q}_i) \cdot (\partial\,\f(\x)/\partial\,\dot{q}_j)$.
We refer to diagonal elements $\Gamma_{ii}$ of the generalized hydrodynamic friction matrix $\boldsymbol{\Gamma}$ as \textit{self-friction} coefficients.
Off-diagonal elements $\Gamma_{ij}$, $i\neq j$, 
or \textit{cross-friction coefficients}, 
characterize a coupling between different degrees of freedom 
(e.g., a coupling between translational and rotational degrees of freedom for chiral objects;
or direct hydrodynamic interactions between different sub-objects that can, in principle, move independently).

The rate of hydrodynamic dissipation can thus be expressed as a quadratic form in the generalized velocity $\dot{\q}$ 
\begin{equation}
\R^{(h)} = \dot{\q}\cdot\boldsymbol{\Gamma}\cdot\dot{\q} = \sum_{i,j} \Gamma_{ij} \, \dot{q}_i\dot{q}_j \quad.
\end{equation}
The hydrodynamic dissipation rate $\R^{(h)}$ 
plays the role of a Rayleigh dissipation function for Lagrangian mechanics of dissipative systems \cite{Goldstein:mechanics}. 
Specifically, we could have equivalently defined the generalized forces as
$2P_i = \partial \R^{(h)} / \partial \dot{q}_i$.
(Following standard notation, the Rayleigh dissipation function is actually $\R^{(h)}/2$ \cite{Goldstein:mechanics}).

The matrix $\boldsymbol{\Gamma}$ is symmetric,
which represents a special case of Onsager reciprocity \cite{Landau:fluid}.
The proof follows directly from the Lorentz reciprocal theorem \cite{Happel:hydro}:
let $\v_i$, $\stress_i$ and $\v_j$, $\stress_j$ denote the flow field and stress tensor 
associated with a change of only $q_i$ with rate $\dot{q}_i$, 
or a change of only $q_j$ with rate $\dot{q}_j$, respectively, 
while all other generalized coordinates are kept constant; 
then
\begin{equation}
\label{eq:gamma_symmetry}
\dot{q}_i\, \Gamma_{ij}(\q)\, \dot{q}_j = 
-\int_\S \! d^2\x\, \v_i\cdot\stress_j\cdot\n \stackrel{(\ast)}{=}
-\int_\S \! d^2\x\, \v_j\cdot\stress_i\cdot\n = 
\dot{q}_j\, \Gamma_{ji}(\q)\, \dot{q}_i 
\quad,
\end{equation}
where we used the Lorentz reciprocal theorem at $(\ast)$.

The matrix $\boldsymbol{\Gamma}$ is also positive semi-definite, consistent with the fact that the rate of energy dissipation should be non-negative.
(In fact, $\boldsymbol{\Gamma}$ should be positive definite, except maybe at singular points $\q$ in configuration space, 
where $\w_i=\partial \x/\partial q_i$, $i=1,\ldots,n$ are linearly dependent.)

In addition to hydrodynamic dissipation as characterized by $\mathcal{R}^{(h)}$, 
internal dissipative processes can be included in our framework, 
provided the corresponding dissipation function is likewise a quadratic form of the generalized velocity \cite{Klindt2016,Klindt2017}.

\section{Equation of motion}

\paragraph{Balance of generalized forces.}

We introduce active driving forces $Q_i$, $i=1,\ldots,n$ 
that coarse-grain internal processes that drive the active shape changes of the active surface.
%
Previous minimal models of flagella synchronization
considered spheres moving along circular orbits driven by a tangential force \cite{Vilfan2006,Niedermayer2008,Uchida2011,Friedrich2012}.
Our active driving forces $Q_i$ generalize the active driving forces considered in these models.

We postulate a balance of generalized forces between driving forces and hydrodynamic friction forces
\begin{equation}
\label{eq:force_balance}
Q_i = P_i,\ i=1,\ldots,n \quad. 
\end{equation}
We emphasize that Eq.~(\ref{eq:force_balance}) is simply an instance of Newton's second law, and thus does not involve any new assumptions.
Simplifying modeling assumptions have only been made in constraining the shape dynamics to a finite number of degrees of freedom $q_1,\ldots,q_n$, 
and in the choice of the active driving forces $Q_1,\ldots,Q_n$. 

From the force balance equation, 
Eq.~(\ref{eq:force_balance}),
and Eq.~(\ref{eq:P_i}) expressing the generalized forces $P_i$,
we obtain equations of motion for the generalized velocities $\dot{\q}$ 
\begin{equation}
\label{eq:motion}
\dot{\q} = \boldsymbol{\Gamma}^{-1}\cdot \Q \quad, 
\end{equation}
where $\Q = (Q_1,\ldots,Q_n)^T$ is the vector of active forces.

\paragraph{Calibration of active driving forces.}
Because each driving force $Q_i$ characterizes internal processes, 
it is plausible to assume that $Q_i$ only depends on the corresponding degree of freedom $q_i$, 
but not on the other $q_j$, $j\neq i$,
i.e., we may assume $Q_i=Q_i(\varphi_i)$.
This assumption will hold in particular in applications, where the index $i$ enumerates different microswimmers or different cilia.  
In principle, $Q_i$ may additionally depend on the friction force $P_i$ itself, 
i.e., if the internal active processes may change under load \cite{Friedrich2018}.
In this case, Eq.~(\ref{eq:force_balance}) becomes a self-consistency equation that has to be solved using methods for implicit equations. 
For a number of biological application cases, 
it was sufficient to assume that $Q_i$ is independent of load \cite{Geyer2013,Klindt2016,Klindt2017}.
In this case, the active driving forces can be uniquely calibrated from a reference dynamics, 
ideally known from experiments. 
Once this is done, one can extrapolate to alternative dynamic scenarios.

As an example for this calibration procedure, 
previous work used experimental data of in-phase synchronized beating in the biflagellate green alga \textit{Chlamydomonas},
which allowed to predict the response to perturbations of this synchronized state \cite{Geyer2013}.
Similarly, measured cilia beat patterns in the absence of external flow have been used
to calibrate active driving forces and predict the response to external flow \cite{Klindt2016}.
In the application section below, 
we consider the dynamics of an isolated cilium with constant phase speed to calibrate its active driving force.
We then use this model to predict synchronization dynamics for a pair of cilia.
In all these cases, the driving forces $Q_i$ coarse-grain internal active processes.

Additionally, 
the formalism allows to incorporate internal elastic degrees of freedom $q_i$ and the corresponding elastic restoring forces $Q_i$
in a formally equivalent manner.
An example includes the waveform compliance of flagellar bending waves \cite{Klindt2016,Klindt2017}.
Similarly, one can include external forces acting on self-propelled shape-changing microswimmers,
as discussed in the next section.

\section{Rigid body motion of a self-propelled microswimmer}

The above formalism includes the important application case of shape-changing microswimmers
and their self-propulsion in a viscous fluid. 
For that aim, we introduce rigid body transformation and include these in the set of generalized coordinates.

Specifically, we consider a microswimmer with outer surface $\S$ 
and introduce a material frame of this microswimmer 
consisting of a reference point $\x_0$ and a set of orthonormal vectors $\e_1$, $\e_2$, $\e_3$.

A rigid body motion of the swimmer is characterized by 
a translation of its reference point, $\dot{\x}_0 = \v_0 = v_1\,\e_1 + v_2\,\e_2 + v_3\,\e_3$,
and a rotation of its material frame with 
$\dot{\e}_i = \varepsilon_{ijk} \Omega_j \e_k$, 
where $\varepsilon_{ijk}$ denotes the Levi-Cevita symbol 
and we use Einstein summation convention.
The components 
$v_1$, $v_2$, $v_3$ and $\Omega_1$, $\Omega_2$ and $\Omega_3$ 
of the translational and the rotational velocity vector with respect to the basis $\e_1$, $\e_2$, $\e_3$, respectively, 
represent the six degrees of freedom of rigid body motion 
and satisfy
$v_1=\v_0\cdot\e_1$, $v_2=\v_0\cdot\e_2$, $v_3=\v_0\cdot\e_3$, and
$\Omega_1=\dot{\e}_2\cdot{\e}_3=-\dot{\e}_3\cdot{\e}_2$,
$\Omega_2=\dot{\e}_3\cdot{\e}_1=-\dot{\e}_1\cdot{\e}_3$,
$\Omega_3=\dot{\e}_1\cdot{\e}_2=-\dot{\e}_2\cdot{\e}_1$.

We choose these velocity components as
the six generalized velocities
\begin{equation}
\label{eq:rigid}
\dot{q}_1 = v_1,\ 
\dot{q}_2 = v_2,\
\dot{q}_3 = v_3,\ 
\dot{q}_4 = \Omega_1,\
\dot{q}_5 = \Omega_2,\
\dot{q}_6 = \Omega_3 \quad.
\end{equation}
The coordinates $q_1, \ldots, q_6$ are elements of the Lie group $\mathfrak{se}(3)=\mathbbm{R}^3\times\mathfrak{so}(3)$ of rigid body transformation
\cite{Murray:rigid}.

For the special case, where the generalized velocities represent rigid body motion as in Eq.~(\ref{eq:rigid}),
the conjugate generalized hydrodynamic friction forces defined in Eq.~(\ref{eq:P_i}) are simply given by the 
components of the total hydrodynamic friction force $\F$ and the total hydrodynamic friction torque $\M$, respectively
\begin{equation}
\label{eq:rigid_F}
P_1 = \F \cdot \e_1,\
P_2 = \F \cdot \e_2,\
P_3 = \F \cdot \e_3,\
P_4 = \M \cdot \e_1,\
P_5 = \M \cdot \e_2,\
P_6 = \M \cdot \e_3 \quad.
\end{equation}
In this case, the $6\times 6$-matrix of generalized hydrodynamic friction coefficients $\boldsymbol{\Gamma}$ 
reduces to the well-known hydrodynamic friction matrix (inverse mobility matrix) of an arbitrary-shaped rigid object.
For a collection of rigid objects (e.g., a collection of rigid spheres as considered in \cite{Reichert:phd}),
we recover the inverse of the grand mobility matrix.

We can describe active shape changes of the microswimmer using coordinates $x'_1$, $x'_2$, $x'_3$ relative to the swimmer's material frame
for each point $\x\in\S$ on the surface
\begin{equation}
\label{eq:xprime}
\x = \x_0 + x'_1\, \e_1 + x'_2\, \e_2 + x'_3\, \e_3 \quad.
\end{equation}
We introduce the time-dependent rigid body transformation 
that maps the material frame of the swimmer to the laboratory frame, 
such that the reference point $\r_0$ of the swimmer is mapped to the origin $\zerovec\in\mathbbm{R}^3$, 
and the material frame vectors are mapped to the standard unit vectors, respectively. 
The coordinates $x'_1$, $x'_2$, $x'_3$ are then just the coordinates of the image $\x'$ of a point $\x\in\S$ under this transformation, 
i.e., the coordinates of the surface after it has been brought into a reference condition \cite{Shapere1987}.
Eq.~(\ref{eq:xprime}) allows us to decompose the displacement velocity $\v(x)$ of the surface into 
a contribution stemming from the rigid body motion and 
a contribution stemming only from any shape change
\begin{equation}
\v(\x) = \dot{\x} = 
\underbrace{
\dot{\x}_0 + \sum_{j=1}^3 x'_j\,\dot{\e}_j 
}_\text{rigid body motion} + 
\underbrace{
\sum_{j=1}^3 \dot{x}'_j\,\e_j 
}_\text{shape change}
\quad.
\end{equation}
The superposition principle of low-Reynolds number flow, Eq.~(\ref{eq:f_superposition}),
implies that the surface density $f(\x)$ of hydrodynamic friction forces can be written as 
a superposition of contributions due to rigid body motion 
and a contribution $\f_\act(\x)$ due the active shape change
\begin{equation}
\label{eq:f_swimmer}
\f(\x) = 
v_1 \,\g_1(\x) + v_2\, \g_2(\x) + v_3\, \g_3(\x) +
\Omega_1\,\g_4(\x) + \Omega_2\,\g_5(\x) + \Omega_3\,\g_6(\x) +
\f_\act(\x)
\quad,
\end{equation}
where
$f_\act(\x)$ depends only on the shape change $\dot{\x}'$, but not on the translational velocity $\v_0$ nor the rotational velocity $\Omegavec$.

Since inertia is assumed negligible, the total force and total torque acting on a microswimmer must equal 
any external force or torque acting on the swimmer, 
$\F = \F^\mathrm{ext}$, $\M=\M^\mathrm{ext}$ \cite{Happel:hydro}.
It follows that a microswimmer that is free from external forces or torques 
does not exert any net force or torque on the surrounding fluid itself
\begin{equation}
\label{eq:balance}
\F = \mathbf{0} \quad,\quad \M = \mathbf{0} \quad.
\end{equation}
Eq.~(\ref{eq:balance}) holds in particular for a neutrally buoyant biological microswimmer (a good approximation for many biological microswimmers).

The surface density of hydrodynamic friction forces due to active shape changes, $\f_\act(\x)$, 
gives rise to a contribution $\F_\act = \int_\S \! d^2\x\, \f_\act(\x)$ to the total force, 
as well as an analogous contribution $\M_\act$ to the total torque. 
The force and torque balance equations, Eq.~(\ref{eq:balance}), 
thus provide an inhomogeneous system of six linear equations
for the six components of the translational and rotational velocity,
$\v_0$ and $\Omegavec$.

We emphasize that Eq.~(\ref{eq:force_balance}) is very general, 
and includes the following application cases of microswimmer motion:
\begin{itemize}
\item
\textit{External forces or torques:}
For example, external forces $\F^\mathrm{ext}$, 
or external torques $\M^\mathrm{ext}$ 
are captured by corresponding external forces $Q_i^\ext$.
Examples include gravitational force for a non-buoyant swimmer or 
torques exerted by an external rotating magnetic fields on an artificial microswimmer with non-zero magnetic dipole moment.
\item
\textit{Prescribed shape dynamics:}
For a prescribed shape-dynamics, say of shape coordinate $q_i$ with prescribed driving protocol $q_i(t)$, 
one would omit the corresponding force balance equation $Q_i=P_i$ from the set of equations Eq.~(\ref{eq:force_balance}), 
and solve for the equation of motion of the other coordinates with prescribed $q_i(t)$.
The conjugate hydrodynamic friction force $P_i$ nonetheless appears in the formula for the total hydrodynamic dissipation rate $\R^{(h)}$, 
where $P_i\dot{q}_i$ equals the rate of work required for the shape change with rate $\dot{q}_i$.
A number of classical theory publications 
on self-propelled biological microswimmers considered prescribed shape dynamics 
\cite{Taylor1951,Gray1955b,Shapere1987,Becker2003,Najafi2004,Dreyfus2005}. 
\item
\textit{Constrained motion.}
Several applications considered constrained swimmers, for example,
biological microswimmers clamped in micropipettes constrained from translational motion \cite{Ruffer1998a,Goldstein2009,Brumley2014}.
Formally, this is a special case of a coordinate $q_i$ with prescribed dynamics
for the coordinates $q_1,\ldots,q_3$ representing rigid body translation, 
enforcing $\dot{q}_i=0$.
The conjugate hydrodynamic friction force $P_i$ equals the external constraining force required to impose the constraint.
Similarly, to constrain a microswimmer from rotational motion 
requires a constraining torque
$\M= P_4\,\e_1 + P_5\,\e_2 + P_6\,\e_3$.
As a historical note, in their classical 1955 paper, 
Gray \& Hancock considered self-propulsion of sperm cells with constrained rotational motion 
to simplify the calculation \cite{Gray1955b}.

Finally, clamped microswimmers exposed to uniform external flow with flow velocity $\u_0$ far from the swimmer as considered in \cite{Klindt2016} 
can be incorporated into our formalism by switching to a co-moving reference frame in which the fluid is at rest.
In the co-moving frame, the clamped swimmer is ``dragged'' through the fluid, 
corresponding to a constraint for rigid body translation,
$\dot{q}_i = -\u_0\cdot\e_i$, $i=1,2,3$. 
Correspondingly, the total hydrodynamic friction force 
$\F= P_1\,\e_1 + P_2\,\e_2 + P_3\,\e_3$
represents the constraining force required to clamp the microswimmer in such an external flow.
\end{itemize}

\section{Multi-scale modeling: Numerical implementation}

To solve for the dynamics of an active surface according to Eq.~(\ref{eq:motion}), 
it suffices to compute the generalized hydrodynamic friction matrix 
$\boldsymbol{\Gamma}$ for a set of reference configurations $\q$ and save this as a look-up table; 
the friction matrix $\boldsymbol{\Gamma}(\q)$ for arbitrary $\q$ can then be found by interpolation.
This allows to solve the equation of motion Eq.~(\ref{eq:motion})
fast, using pre-computed hydrodynamic friction coefficients. 
We outline the numerical implementation of this general program.

While Eq.~(\ref{eq:Gamma}) may look abstract, all quantities can be directly obtained from numerical computations for arbitrary surfaces $\S$.
Assume the surface $\S$ is represented by a triangulated mesh.
The triangular faces (or `elements') shall be enumerated by $k\in\I$ 
with midpoints $\x_k$
and respective areas $A_k$.

In a first step, 
we compute a (normalized) surface distribution of velocities
$\w_i(\x_k)$, $k\in\I$ 
for each generalized coordinate $i=1,\ldots,n$, 
either by computing the derivative
$\w_i(\x_k) = \partial \x_k(\q) / \partial q_i$ 
analytically,
or by evaluating the finite difference quotient
\begin{equation}
\w_i(\x_k) = \frac{ \x_k (\q + \varepsilon\,\Delta_i) - \x_k(\q) }{\varepsilon} \comma
\end{equation}
for each midpoint $\x_k$, $k\in\I$,
where 
$\Delta_i$ is the unit vector whose components are all zero, except the $i^\mathrm{th}$ component,
and $\varepsilon$ is a small number.

We can use boundary element methods
to numerically compute a surface density of hydrodynamic friction forces
$\f(\x_k)$
with physical units of a stress, 
given an arbitrary surface distribution of velocities
$\v(\x_k)$ specified at each midpoint $\x_k$, $k\in\I$.
Specifically,
in the application example below, we use
the fast multi-pole boundary element method \textit{fastBEM} \cite{Liu2006,Liu2009}. 

In the next step, 
we compute the surface density $\f_j(\x_k) = \alpha_j\,\g_j(\x_k)$ of hydrodynamic friction forces,
corresponding to the velocity distribution
$\v_j(\x_k)=\alpha_j\,\w_j(\x_k)$.
Here, $\alpha_j$ is an arbitrary constant to ensure proper physical units of a velocity for $\v_j$.
We thus obtain $n$ surface distributions of (normalized)
hydrodynamic friction forces 
$\g_j(\x_k)$, $j=1,\ldots,n$,
one for each generalized coordinate $q_j$. 
These
 force distributions $\g_j(\x_k)$ depend on $\q$, but not on $\dot{\q}$.
Finally, we compute the components $\Gamma_{ij}$ of the generalized hydrodynamic friction matrix $\boldsymbol{\Gamma}$
by taking the scalar product of 
the $i^\mathrm{th}$ (normalized) velocity distribution $\w_i(\x_k)$, and 
the $j^\mathrm{th}$ (normalized) force distribution $\g_j(\x_k)$
\begin{equation}
\Gamma_{ij} =  \sum_{k\in\I} \w_i(\x_k)\cdot \g_j(\x_k)\, A_k
\comma
i,j=1,\ldots,n
\quad,
\end{equation}
where $A_k$ was the area of the $k^\mathrm{th}$ triangle.
We can interpret $\alpha_j\g_j(\x_k)A_k$ at the total force acting on the $k^\mathrm{th}$ element
(with proper physical units of a force)
if the generalized coordinate $q_i$ would change at a rate $\alpha_i$.

Importantly, it suffices to compute the generalized hydrodynamic friction matrix 
$\boldsymbol{\Gamma}$ only for a set of reference configurations and save this as a look-up table. 
If $m$ discrete values are used for each of the $n$ generalized coordinates,
the Stokes equation needs to be solved
a total of $n\,m^n$ times,
as we need to change each of the $q_j$, $j=1,\ldots,n$ for $m^n$ different choices of $\q$.
By exploiting symmetries,
as well as translational and rotational invariance for individual microswimmers,
this number can be reduced further.
The friction matrix $\boldsymbol{\Gamma}(\q)$ for arbitrary $\q$ can then be found by interpolation. 
For example, spline interpolation, low-order polynomials, and (double) Fourier series 
were used in previous applications \cite{Geyer2013,Klindt2016,Klindt2017}.

In principle, different hydrodynamic simulation methods could be used to solve the Stokes equation and compute the force distribution $\f(\x_k)$.
Deterministic lattice Boltzmann solvers may be suitable, provided the effective Reynolds numbers are sufficiently small.
An early application represented the surface of a microswimmer not by a triangulated mesh, 
but as a collection of equally-sized spheres, 
and computed the grand mobility matrix for these spheres using the \textit{hydrolib} package \cite{Hinsen1995}.
In the application example below, we employ the fast multi-pole boundary element method \textit{fastBEM} \cite{Liu2006,Liu2009}, 
available for download at \cite{Liu:url}.
The open source implementation of the fast boundary element method \textit{STKFMM} 
directly incorporates the fundamental solution of the Stokes equation close to a no-slip boundary wall~\cite{Blake1971}, 
and thus relieves the need for an explicit representation of the boundary as a triangulated mesh,
yet currently only supports the computation of velocity fields from force distributions \cite{Yan2018, Yan:url}.

\section{Application: pair of interacting cilia}

We demonstrate our LAMAS modeling framework using the example of hydrodynamic synchronization in pairs of cilia. 
We thereby reconsider the question of in-phase and anti-phase synchronization previously addressed by Vilfan et al. \cite{Vilfan2006}, 
yet, instead of a minimal model of spheres orbiting on circular trajectories, 
we employ in our simulations a realistic cilia beat pattern obtained from previous experiments. 

\begin{figure}
\includegraphics[width=0.7\linewidth]{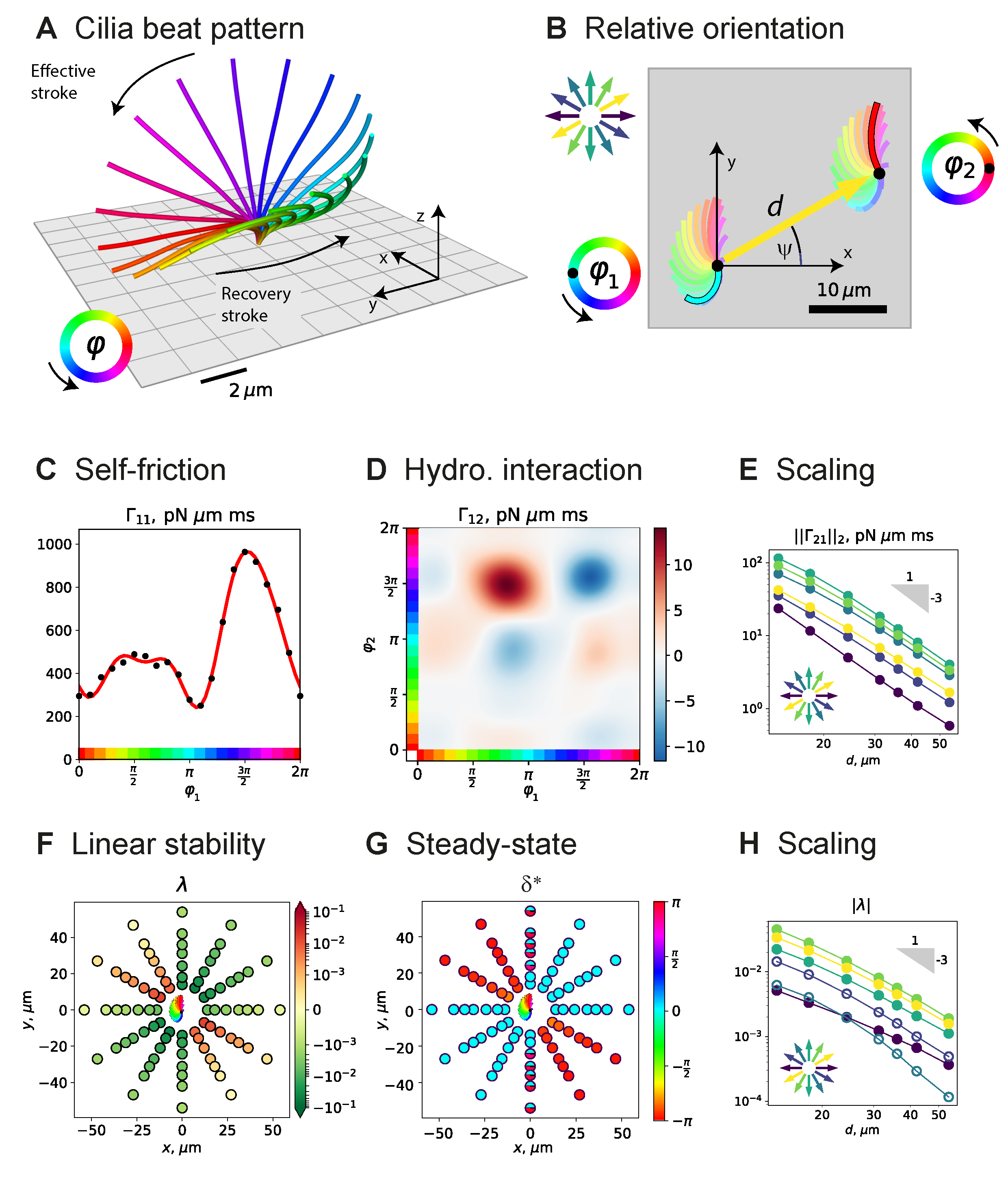}
\caption{
\textbf{In-phase and anti-phase synchronization in a pair of interacting cilia.}
(A)
Cilia beat pattern from unicellular \textit{Paramecium} \cite{Machemer1972} as reported in \cite{Naitoh1984},
shown as sequence of three-dimensional shapes parameterized by a $2\pi$-periodic phase variable $\varphi$ (color code).
Spacing of square grid: $2\,\micron$.
(B)
We consider a pair of cilia with respective phases $\varphi_1$ and $\varphi_2$,
whose base points are separated by a distance $d$
along a direction that encloses an angle $\psi$ with the $x$ axis 
(where the $y$ axis is set by the direction of the effective stroke of both cilia).
(C)
Self-friction coefficient $\Gamma_{11}(\varphi_1)$ of a single cilium as function of its phase variable $\varphi_1$,
obtained by solving the Stokes equation of three-dimensional flow (blue dots),
as well as continuous representation as Fourier series (orange line).
In the case of a single cilium,
$\Gamma_{11}$ is proportional to the phase-dependent active cilia driving force $Q_1(\varphi_1)$.
(D)
Generalized hydrodynamic
friction coefficient $\Gamma_{12}(\varphi_1,\varphi_2)$ 
characterizing hydrodynamic interactions from the second cilium to the first cilium,
see also Eq.~(\ref{eq:cilia_pair}).
Positive values (red colors) imply that the motion of the second cilium causes the first cilium to beat slower,
while negative values (blue colors) imply that the first cilium beats faster.
Cilia distance $d = 18\,\micron$, orientation angle $\psi = 2\pi/3$. 
(E)
The magnitude of hydrodynamic interactions,
here quantified by the $L_2$-norm of $\Gamma_{12}$,
decay as $\sim 1/d^3$,
consistent with the theoretical scaling expected from the Blake tensor \cite{Blake1971}.
Different curves correspond to different separation directions between the two cilia
($\psi=0$: dark-blue, $\psi=\pi/3$: light-green, $\psi=2\pi/3$: teal; also indicated by the direction arrows.)
(F)
We characterize the stability of the in-phase synchronized state,
defined by $\varphi_1(t)=\varphi_2(t)$,
for different relative orientations of the two cilia by a Lyapunov exponent $\lambda$, see Eq.~(\ref{eq:lambda}).
Colored dots at respective positions in the $xy$-plane represent the value of $\lambda$
if the second cilium is positioned at the position of the dot and the first cilium is located at the origin.
Negative values imply that in-phase synchronization is linearly stable (green colors, $\lambda<0$),
while positive values imply that in-phase synchronization is linearly unstable (red colors, $\lambda>0$).
(G)
We determined the steady-state phase difference $\delta^\ast$ between the two cilia for different relative cilia positions,
analogous to panel F.
While $\delta^\ast=0$ for cilia orientations with stable in-phase synchronization (cyan),
we observe anti-phase synchronization with $\delta^\ast\approx\pi$ for cilia orientations with $\lambda>0$ (red colors).
For relative cilia orientation aligned with the direction of the effective stroke of the cilia beat 
($\psi=\pi/2$),
we observed cases of multi-stability (bi-colored dots).
(H)
Consistent with the far-field scaling of hydrodynamic interactions as shown in panel E,
we find that also the Lyapunov exponent $\lambda$, which represents an effective synchronization strength,
likewise decays as $1/d^3$ with distance $d$ between the two cilia.
Different curves represent different separation directions, analogous to panel E.
Frequency of cilia beat: $\omega_0 /(2\pi) = 32 \,\mathrm{Hz}$ \cite{Machemer1972},
fluid viscosity: $\mu = 10^{-3}\, \textrm{Pa}\,\textrm{s}$.
}
\label{figure3}
\end{figure}

We digitalized and reconstructed three-dimensional shapes of a beating cilium 
on the surface of the unicellular ciliated protist \textit{Paramecium} \cite{Machemer1972} 
as presented in \cite{Naitoh1984}.
The cilia beat is periodic, 
and we can thus describe the shape of the cilia centerline 
as a periodic shape sequence parametrized by a $2\pi$-periodic phase variable $\varphi$, 
see Fig.~\ref{figure3}A.
For unperturbed beating, the phase speed equals the angular frequency of the cilia beat,
$\dot \varphi (t) = \omega_0$.

\subsection{Equation of motion for a pair of cilia}

We consider two identical cilia beating in the same direction attached to a no-slip boundary wall, 
see Fig.~\ref{figure3}A and B. 
We describe each cilium by a single phase variable that parameterizes its periodic sequence of centerline shapes.
The two phase variables $\varphi_1$ and $\varphi_2$ fully characterize the dynamics of the two beating cilia, 
and represent a set of generalized coordinates with state vector $\q=(\varphi_1,\varphi_2)$.

For our example, the force balance equation, Eq.~(\ref{eq:force_balance}) takes the form
\begin{align}
\label{eq:cilia_pair}
Q_1(\varphi_1) &= \Gamma_{11}(\varphi_1, \varphi_2) \dot{\varphi}_1 + \Gamma_{12}(\varphi_1, \varphi_2) \dot{\varphi}_2 \notag \\
Q_2(\varphi_2) &= \Gamma_{21}(\varphi_1, \varphi_2) \dot{\varphi}_1 + \Gamma_{22}(\varphi_2, \varphi_2) \dot{\varphi}_2 \quad.
\end{align}
This equation can be further simplified.
The symmetry relation Eq.~\eqref{eq:gamma_symmetry} implies
$\Gamma_{12}(\varphi_1,\varphi_2) = \Gamma_{21}(\varphi_1,\varphi_2)$.
Numerical computation of 
$\Gamma_{11}(\varphi_1, \varphi_2)$ 
shows that this self-friction coefficient of the first cilium 
is virtually independent of the phase of the second cilium, 
and almost does not change when the other cilium is not present at all.
An analogous statement holds for the second cilium.
Therefore, we can replace the two self-friction coefficients in Eq.~(\ref{eq:cilia_pair}), 
$\Gamma_{11}(\varphi_1, \varphi_2)$ and $\Gamma_{22}(\varphi_1, \varphi_2)$,
by the self-friction coefficient
for a single cilium to very good approximation.
This approximation allows us to define the active driving forces using the case of a single cilium.

\paragraph{Calibration of active driving force.}
We require that a single cilium 
should beat at a constant phase speed
$\dot{\varphi}_1 = \omega_0$,
where $\omega_0$ denotes the intrinsic beat frequency of the cilium
if there are no interactions with other cilia.
This requirement uniquely determines the active driving force $Q_1(\varphi_1)$.
Specifically, 
for a single cilium, 
the force balance equation reads,
$Q_1(\varphi_1) = \Gamma_{11}(\varphi_1)\,\dot{\varphi}_1$.
We conclude
$Q_1(\varphi_1) = \omega_0\, \Gamma_{11}(\varphi_1)$;
Fig.~\ref{figure3}C displays the phase-dependence of $\Gamma_{11}(\varphi_1)$.
Since both cilia are assumed identical with same intrinsic beat frequency $\omega_0$, 
this also specifies the active driving force $Q_2(\varphi_2)$ of the second cilium.


\paragraph{Equation of motion.}
Using the force balance equation, Eq.~\ref{eq:cilia_pair}, 
and the calibrated driving force, 
we obtain the equation of motion
\begin{align}
\label{eq:cilia_pair_motion}
\dot{\varphi_1} &= \omega_0 - C_1(\varphi_1,\varphi_2) \,\dot{\varphi}_2 
\quad,\quad
C_1(\varphi_1,\varphi_2) = \frac{\Gamma_{12}(\varphi_1, \varphi_2)}{\Gamma_{11}(\varphi_1)} \notag \\
\dot{\varphi_2} &= \omega_0 - C_2(\varphi_1,\varphi_2) \,\dot{\varphi}_1
\quad,\quad
C_2(\varphi_1,\varphi_2) = \frac{\Gamma_{12}(\varphi_1, \varphi_2)}{\Gamma_{11}(\varphi_2)} \quad.
\end{align} 
Eq.~(\ref{eq:cilia_pair_motion}) describes a pair of coupled phase oscillators.

In the following, 
we use Eq.~(\ref{eq:cilia_pair_motion}) and pre-computed friction coefficients to analyze in-phase and anti-phase synchronization
of the two cilia depending on their relative position. 
Details on the numerical computation of 
$\Gamma_{ij}(\varphi_1,\varphi_2)$ can be found in the appendix.
An example of the generalized friction coefficient $\Gamma_{12}(\varphi_1, \varphi_2)$,
which characterizes hydrodynamic interactions between the two cilia,
is shown in Fig.~\ref{figure3}D; Fig.~\ref{figureS2} shows $\Gamma_{12}(\varphi_1,\varphi_2)$ for additional cilia orientations.

\subsection{Results: In-phase and anti-phase synchronization as function of direction}

\paragraph{Hydrodynamic interactions decay as $1/d^3$.}

For large separation distances $d$ between the two cilia, 
hydrodynamic interactions between the two cilia as characterized by $\Gamma_{12}(\varphi_1,\varphi_2)$ decay as $1/d^3$, see Fig.~\ref{figure3}E.
This asymptotic scaling is consistent with the expected leading order singularity of the flow field induced by a single cilium.
Specifically, the flow field induced by a point force parallel to a no-slip boundary wall is given by
the Blake tensor \cite{Blake1971},
and decays as $1/d^3$ for points at a constant height from the boundary,
which is the relevant case for the interaction between cilia \cite{Guirao2007}.

\paragraph{Linear stability analysis.}

Since both cilia were assumed identical,
the in-phase synchronized state with $\varphi_1(t)=\varphi_2(t)$
is always a periodic solution of Eq.~(\ref{eq:cilia_pair_motion}).
To assess the linear stability of this in-phase synchronized state,
we monitored the evolution of a small perturbation 
of the phase difference 
$\delta(t) = \varphi_2(t)-\varphi_1(t)$
during one beat cycle. 
Specifically, 
we integrated Eq.~(\ref{eq:cilia_pair_motion}) with the initial condition
$\varphi_1(t=0)=-\delta_0/2$ and $\varphi_2(t=0)=\delta_0/2$ 
for a small perturbation $|\delta_0|\ll 1$
up to time $T$ defined by $\ol{\varphi}(T) = [\varphi_1(T)+\varphi_2(T)]/2=2\pi$
(corresponding to the completion of a full beat cycle), 
and recorded the new phase difference
$\delta_1=\delta(T)$.

We define a dimensionless Lyapunov exponent as
\begin{equation}
\label{eq:lambda}
\lambda = \log  |\delta_1 / \delta_0|\quad,
\end{equation}
which characterizes whether the initial perturbation decays or grows.
The in-phase synchronized state is linearly stable if 
$|\delta_1| < |\delta_0|$ (hence $\lambda < 0$), 
and linearly unstable if 
$|\delta_1| > |\delta_0|$ (hence $\lambda > 0$).

Fig.~\ref{figure3}F shows $\lambda$ as function of relative cilia position.
Here, the first cilium is located at the origin, 
while the second cilium is located at the position of the respective colored dots.

The symmetry of Eq.~(\ref{eq:cilia_pair_motion})
implies that the synchronization behavior is invariant under a point reflection, which swaps cilia $1$ and $2$. 
Whether in-phase synchronization is stable or not 
only depends on the direction of the separation vector between the two cilia
(where $\lambda>0$ for direction angles $\psi=2\pi/3$ and $5\pi/6$, in which case the cilia synchronize anti-phase, 
as discussed next).

Additionally, we analyzed the steady-state dynamics of Eq.~(\ref{eq:cilia_pair_motion})
and identified phase differences $\delta^\ast$ that correspond to stable periodic solutions, see Fig.~\ref{figure3}G.
As a technical point, $\delta(t)$ may weakly oscillate during each cycle; 
we therefore define $\delta^\ast$ as the phase difference at times for which $\ol{\varphi}$ is an integer multiple of $2\pi$.

When the in-phase synchronized state is linearly stable for a given cilia configuration, 
we obviously have $\delta^\ast=0$.
If, however, the in-phase synchronized state is linearly unstable, we approximately find
$\delta^\ast\approx\pi$, 
corresponding to \textit{anti-phase synchronization}.
For a few cilia configurations, we observe multi-stability, 
characterized by two different values of the phase differences $\delta^\ast$ that correspond to stable periodic solutions;
these configurations are indicated as bi-colored half circles in Fig.~\ref{figure3}G.

The magnitude $|\lambda|$ of the Lyapunov exponents decreases as $1/d^3$ with distance $d$ between the two cilia, see Fig.~\ref{figure3}H, 
consistent with the far-field scaling of hydrodynamic interactions shown in panel E.
This suggests that short-range interactions between close-by cilia may dominate emergent behavior in carpets of many cilia.

\section{Discussion}

\paragraph{Summary.}
We presented a multi-scale modeling and simulation framework for active surfaces immersed in viscous fluids.
This includes self-propulsion of shape-changing microswimmers as a special case.
The key idea is to constrain the shape dynamics to a small number of principal deformation modes.
These modes represent generalized coordinates, 
for which generalized hydrodynamic friction coefficients are defined according to the formalism of Lagrangian mechanics.
To actually compute these friction coefficients, 
the Stokes equation is solved for an infinitesimal change of each generalized coordinate in an initial step.
For subsequent dynamic simulations, 
a generalized force balance between hydrodynamic friction forces and 
active driving forces is solved in each time step. 
This is sufficiently fast since this second step does not involve any hydrodynamic computations, 
but uses the pre-computed hydrodynamic friction coefficients.

Our formalism generalizes classical Lagrangian dynamics of dissipative systems \cite{Goldstein:mechanics} to active systems. 
The rate of work exerted by the active surface on the surrounding fluid provides a Rayleigh dissipation function $\R^{(h)}$, 
which defines generalized friction forces $P_i$ conjugate to each generalized coordinate $q_i$
as a partial derivative $2P_i = \partial \R^{(h)}/\partial q_i$.
Numerically, the generalized friction forces are computed from a surface density of hydrodynamic friction forces using a Lagrangian projection method.
Active driving forces coarse-grain internal active processes, such as the dynamics of molecular motors inside cilia and flagella.
These active driving forces can be calibrated from a reference data set, 
for which the dynamics is already known or prescribed.

Our approach shares the idea of multi-scale modeling to efficiently explore biological fluid dynamics problems at low Reynolds numbers
with recent developments of reduced-order models,
which likewise propose a decomposition of biological hydrodynamics problems with multiple queries
into an initial setup phase during which the Stokes equation needs to be solved for example configurations (`offline phase'), 
and an subsequent phase of parameter space exploration (`online phase') \cite{Giuliani2020}. 
However, our approach does not require an affine dependence of hydrodynamic quantities on model parameters.

\paragraph{Potential applications.}
We applied our general framework to a model example of mutual synchronization between two cilia, 
using an experimentally measured cilia beat pattern.
Future work will generalize this approach to cilia carpets with many cilia,
which previously had been either studied using detailed simulations with many degrees of freedom \cite{Gueron1999,Elgeti2013},
or using minimal models \cite{Guirao2007,Wollin2011,Kotar2013,Meng2020}. 
A key simplifying assumption will be to approximate many-body hydrodynamic interactions between many cilia 
as a superposition of pairwise interactions. 
A similar approach can be applied to study self-organized pattern formation in suspension of shape-changing microswimmers, 
using the approximation of pairwise interactions between microswimmers, which is valid for dilute suspensions.

An important feature of our modeling framework is that \emph{biological noise} can be incorporated in a natural way. 
Beating cilia are known to exhibit both phase fluctuations (frequency jitter), 
as well as amplitude fluctuations \cite{Polin2009,Goldstein2009,Ma2014}.
This active noise jeopardizes synchronization of cilia by hydrodynamic interactions.
Additionally, noise randomizes the motion of biological microswimmers.
While thermal noise causes noticeable rotational diffusional of micrometer-sized bacteria such as \textit{E. coli} \cite{Berg:ecoli},
amplitude fluctuations of cilia beating affect the swimming of ten-fold larger eukaryotic swimmers \cite{Klindt2015}.
In our framework, active noise is incorporated by using stochastic active driving forces.
In previous work, adding additive Gaussian white noise with noise strengths calibrated from experiments
was sufficient to account for effective diffusion of swimming sperm cells, or noisy synchronization of coupled cilia \cite{Ma2014}.
For simulations accounting for biological noise, 
it is beneficial to use a deterministic solver for the Stokes equation as done here, 
in order to not confound physically relevant noise and fluctuations from a stochastic hydrodynamic simulation method.

Next, our modeling framework helps to conceptualize the \emph{load-response of cilia and flagella}, 
which beat slower if the hydrodynamic load opposing their beat increases.
Classical work showed how cilia and flagella beat slower at increased viscosity of the surrounding fluid \cite{Brokaw1966, Machemer1972};
likewise external flows change the speed of cilia beating \cite{Klindt2016,Friedrich2018,Pellicciotta2020}.
The load response of cilia and flagella is a prerequisite 
for putative mechanisms of synchronization by hydrodynamic or mechanical interactions.
We propose that the generalized hydrodynamic friction force defined here
can serve as a proxy for the effective hydrodynamic load acting on an actively shape-changing structure
such as a beating cilium.

\paragraph{Limitations.}
Our approach is restricted to the limit of zero Reynolds numbers, 
because it crucially relies on the superposition principle for Stokes flow. 
In a laminar flow regime at finite Reynolds numbers, 
we expect that computations of self-friction are still accurate, 
but long-range hydrodynamic interactions become increasingly less accurate with increasing distance
if the Stokes equation is used.
Nonetheless, our approach should still serve as a reasonable approximation, 
since any long-range hydrodynamic interactions that are incorrectly predicted by the Stokes equation
will be very weak already. 

In principle, a similar framework could be developed using the linearized Navier-Stokes equations instead of the Stokes equation used here, 
but only in Fourier space.
The linearized Navier-Stokes equation provides a refined approximation for long-ranged hydrodynamic interactions
if the Reynolds number for oscillatory motion becomes appreciable.
In this case, a superposition principle applies for time-periodic flows.
However, working in frequency space instead of the time domain will make the practical solution of dynamic problems more difficult.

Another limitation of our approach is that it is inherently restricted to Newtonian fluids. 
While certain important biological fluid dynamics problems involve visco-elastic fluids, 
the lack of a superposition principle in this case implies that other methods need to be used.

\paragraph{Conclusion.}
Our modeling and simulation framework LAMAS can be complimentary to existing methods.
Our approach is particularly suited 
to screen extensive parameter ranges, 
provided the modified parameters concern the dynamical model 
(such as active driving forces or effective elastic constants \cite{Klindt2017}),
and do not require re-computation of the generalized hydrodynamic friction coefficients. 
Likewise, our approach allows to compute multiple stochastic realizations of the same problem fast.\\[1mm]

\begin{acknowledgments}
\textit{Acknowledgments.}
AS and BMF are supported by the German National Science Foundation (DFG) 
through the \textit{Microswimmers} priority program (DFG grant FR3429/1-1 and FR3429/1-2 to BMF), 
as well as 
through the Excellence Initiative by the German Federal and State Governments 
(Clusters of Excellence cfaed EXC-1056 and PoL EXC-2068). 
BMF acknowledges a Heisenberg grant (DFG grant FR3429/4-1).

We thank Andrej Vilfan for fruitful discussions.
\end{acknowledgments}

\bibliography{cilia_carpet}

\begin{thebibliography}{76}
\expandafter\ifx\csname natexlab\endcsname\relax\def\natexlab#1{#1}\fi
\expandafter\ifx\csname bibnamefont\endcsname\relax
  \def\bibnamefont#1{#1}\fi
\expandafter\ifx\csname bibfnamefont\endcsname\relax
  \def\bibfnamefont#1{#1}\fi
\expandafter\ifx\csname citenamefont\endcsname\relax
  \def\citenamefont#1{#1}\fi
\expandafter\ifx\csname url\endcsname\relax
  \def\url#1{\texttt{#1}}\fi
\expandafter\ifx\csname urlprefix\endcsname\relax\def\urlprefix{URL }\fi
\providecommand{\bibinfo}[2]{#2}
\providecommand{\eprint}[2][]{\url{#2}}

\bibitem[{\citenamefont{Berg and Anderson}(1973)}]{Berg1973}
\bibinfo{author}{\bibfnamefont{H.~C.} \bibnamefont{Berg}} \bibnamefont{and}
  \bibinfo{author}{\bibfnamefont{R.~A.} \bibnamefont{Anderson}},
  \bibinfo{journal}{Nature} \textbf{\bibinfo{volume}{245}},
  \bibinfo{pages}{380} (\bibinfo{year}{1973}).

\bibitem[{\citenamefont{Wada and Netz}(2007)}]{Wada2007}
\bibinfo{author}{\bibfnamefont{H.}~\bibnamefont{Wada}} \bibnamefont{and}
  \bibinfo{author}{\bibfnamefont{R.~R.} \bibnamefont{Netz}},
  \bibinfo{journal}{Phys. Rev. Lett.} \textbf{\bibinfo{volume}{99}},
  \bibinfo{pages}{108102} (\bibinfo{year}{2007}).

\bibitem[{\citenamefont{Gray}(1928)}]{Gray1928}
\bibinfo{author}{\bibfnamefont{J.}~\bibnamefont{Gray}},
  \emph{\bibinfo{title}{{Ciliary Movements}}} (\bibinfo{publisher}{Cambridge
  Univ. Press}, \bibinfo{address}{Cambridge}, \bibinfo{year}{1928}).

\bibitem[{\citenamefont{Gray}(1955)}]{Gray1955a}
\bibinfo{author}{\bibfnamefont{J.}~\bibnamefont{Gray}}, \bibinfo{journal}{J.
  exp. Biol.} \textbf{\bibinfo{volume}{32}}, \bibinfo{pages}{775}
  (\bibinfo{year}{1955}).

\bibitem[{\citenamefont{Sanderson and Sleigh}(1981)}]{Sanderson1981}
\bibinfo{author}{\bibfnamefont{M.}~\bibnamefont{Sanderson}} \bibnamefont{and}
  \bibinfo{author}{\bibfnamefont{M.}~\bibnamefont{Sleigh}},
  \bibinfo{journal}{J. Cell Sci.} \textbf{\bibinfo{volume}{47}},
  \bibinfo{pages}{331} (\bibinfo{year}{1981}).

\bibitem[{\citenamefont{Faubel et~al.}(2016)\citenamefont{Faubel, Westendorf,
  Bodenschatz, and Eichele}}]{Faubel2016}
\bibinfo{author}{\bibfnamefont{R.}~\bibnamefont{Faubel}},
  \bibinfo{author}{\bibfnamefont{C.}~\bibnamefont{Westendorf}},
  \bibinfo{author}{\bibfnamefont{E.}~\bibnamefont{Bodenschatz}},
  \bibnamefont{and} \bibinfo{author}{\bibfnamefont{G.}~\bibnamefont{Eichele}},
  \bibinfo{journal}{Science} \textbf{\bibinfo{volume}{353}},
  \bibinfo{pages}{176} (\bibinfo{year}{2016}).

\bibitem[{\citenamefont{Brokaw}(1972)}]{Brokaw1972}
\bibinfo{author}{\bibfnamefont{C.~J.} \bibnamefont{Brokaw}},
  \bibinfo{journal}{Biophys. J.} \textbf{\bibinfo{volume}{12}},
  \bibinfo{pages}{564} (\bibinfo{year}{1972}).

\bibitem[{\citenamefont{Lindemann}(1994)}]{Lindemann1994}
\bibinfo{author}{\bibfnamefont{C.~B.} \bibnamefont{Lindemann}},
  \bibinfo{journal}{J. Theoret. Biol.} \textbf{\bibinfo{volume}{168}},
  \bibinfo{pages}{175} (\bibinfo{year}{1994}).

\bibitem[{\citenamefont{Riedel-Kruse et~al.}(2007)\citenamefont{Riedel-Kruse,
  Hilfinger, Howard, and J{\"u}licher}}]{Riedel2007}
\bibinfo{author}{\bibfnamefont{I.~H.} \bibnamefont{Riedel-Kruse}},
  \bibinfo{author}{\bibfnamefont{A.}~\bibnamefont{Hilfinger}},
  \bibinfo{author}{\bibfnamefont{J.}~\bibnamefont{Howard}}, \bibnamefont{and}
  \bibinfo{author}{\bibfnamefont{F.}~\bibnamefont{J{\"u}licher}},
  \bibinfo{journal}{HFSP J.} \textbf{\bibinfo{volume}{1}}, \bibinfo{pages}{192}
  (\bibinfo{year}{2007}).

\bibitem[{\citenamefont{Klindt et~al.}(2016)\citenamefont{Klindt, Ruloff,
  Wagner, and Friedrich}}]{Klindt2016}
\bibinfo{author}{\bibfnamefont{G.~S.} \bibnamefont{Klindt}},
  \bibinfo{author}{\bibfnamefont{C.}~\bibnamefont{Ruloff}},
  \bibinfo{author}{\bibfnamefont{C.}~\bibnamefont{Wagner}}, \bibnamefont{and}
  \bibinfo{author}{\bibfnamefont{B.~M.} \bibnamefont{Friedrich}},
  \bibinfo{journal}{Phys. Rev. Lett.} \textbf{\bibinfo{volume}{117}},
  \bibinfo{pages}{258101} (\bibinfo{year}{2016}).

\bibitem[{\citenamefont{Riedel et~al.}(2005)\citenamefont{Riedel, Kruse, and
  Howard}}]{Riedel2005}
\bibinfo{author}{\bibfnamefont{I.~H.} \bibnamefont{Riedel}},
  \bibinfo{author}{\bibfnamefont{K.}~\bibnamefont{Kruse}}, \bibnamefont{and}
  \bibinfo{author}{\bibfnamefont{J.}~\bibnamefont{Howard}},
  \bibinfo{journal}{Science} \textbf{\bibinfo{volume}{309}},
  \bibinfo{pages}{300} (\bibinfo{year}{2005}).

\bibitem[{\citenamefont{Machemer}(1972)}]{Machemer1972}
\bibinfo{author}{\bibfnamefont{H.}~\bibnamefont{Machemer}},
  \bibinfo{journal}{J. exp. Biol.} \textbf{\bibinfo{volume}{57}},
  \bibinfo{pages}{239} (\bibinfo{year}{1972}).

\bibitem[{\citenamefont{Berg and Turner}(1993)}]{Berg1993}
\bibinfo{author}{\bibfnamefont{H.~C.} \bibnamefont{Berg}} \bibnamefont{and}
  \bibinfo{author}{\bibfnamefont{L.}~\bibnamefont{Turner}},
  \bibinfo{journal}{Biophys. J.} \textbf{\bibinfo{volume}{65}},
  \bibinfo{pages}{2201} (\bibinfo{year}{1993}).

\bibitem[{\citenamefont{Brokaw}(1966)}]{Brokaw1966}
\bibinfo{author}{\bibfnamefont{C.~J.} \bibnamefont{Brokaw}},
  \bibinfo{journal}{J. exp. Biol.} \textbf{\bibinfo{volume}{45}},
  \bibinfo{pages}{113} (\bibinfo{year}{1966}).

\bibitem[{\citenamefont{Golestanian and Ajdari}(2008)}]{Golestanian2008}
\bibinfo{author}{\bibfnamefont{R.}~\bibnamefont{Golestanian}} \bibnamefont{and}
  \bibinfo{author}{\bibfnamefont{A.}~\bibnamefont{Ajdari}},
  \bibinfo{journal}{Phys. Rev. Lett.} \textbf{\bibinfo{volume}{100}},
  \bibinfo{pages}{038101} (\bibinfo{year}{2008}).

\bibitem[{\citenamefont{Pickl et~al.}(2017)\citenamefont{Pickl, Pande,
  K{\"o}stler, R{\"u}de, and Smith}}]{Pickl2017}
\bibinfo{author}{\bibfnamefont{K.}~\bibnamefont{Pickl}},
  \bibinfo{author}{\bibfnamefont{J.}~\bibnamefont{Pande}},
  \bibinfo{author}{\bibfnamefont{H.}~\bibnamefont{K{\"o}stler}},
  \bibinfo{author}{\bibfnamefont{U.}~\bibnamefont{R{\"u}de}}, \bibnamefont{and}
  \bibinfo{author}{\bibfnamefont{A.-S.} \bibnamefont{Smith}},
  \bibinfo{journal}{Journal of Physics: Condensed Matter}
  \textbf{\bibinfo{volume}{29}}, \bibinfo{pages}{124001}
  (\bibinfo{year}{2017}).

\bibitem[{\citenamefont{Friedrich}(2018)}]{Friedrich2018}
\bibinfo{author}{\bibfnamefont{B.~M.} \bibnamefont{Friedrich}},
  \bibinfo{journal}{Phys. Rev. E} \textbf{\bibinfo{volume}{97}},
  \bibinfo{pages}{042416} (\bibinfo{year}{2018}).

\bibitem[{\citenamefont{Wensink et~al.}(2012)\citenamefont{Wensink, Dunkel,
  Heidenreich, Drescher, Goldstein, L{\"o}wen, and Yeomans}}]{Wensink2012}
\bibinfo{author}{\bibfnamefont{H.~H.} \bibnamefont{Wensink}},
  \bibinfo{author}{\bibfnamefont{J.}~\bibnamefont{Dunkel}},
  \bibinfo{author}{\bibfnamefont{S.}~\bibnamefont{Heidenreich}},
  \bibinfo{author}{\bibfnamefont{K.}~\bibnamefont{Drescher}},
  \bibinfo{author}{\bibfnamefont{R.~E.} \bibnamefont{Goldstein}},
  \bibinfo{author}{\bibfnamefont{H.}~\bibnamefont{L{\"o}wen}},
  \bibnamefont{and} \bibinfo{author}{\bibfnamefont{J.~M.}
  \bibnamefont{Yeomans}}, \bibinfo{journal}{Proc. Natl. Acad. Sci. U.S.A.}
  \textbf{\bibinfo{volume}{109}}, \bibinfo{pages}{14308}
  (\bibinfo{year}{2012}).

\bibitem[{\citenamefont{R{\"{u}}ffer and Nultsch}(1998)}]{Ruffer1998a}
\bibinfo{author}{\bibfnamefont{U.}~\bibnamefont{R{\"{u}}ffer}}
  \bibnamefont{and} \bibinfo{author}{\bibfnamefont{W.}~\bibnamefont{Nultsch}},
  \bibinfo{journal}{Cell Motil. Cytoskel.} \textbf{\bibinfo{volume}{41}},
  \bibinfo{pages}{297} (\bibinfo{year}{1998}).

\bibitem[{\citenamefont{Goldstein et~al.}(2009)\citenamefont{Goldstein, Polin,
  and Tuval}}]{Goldstein2009}
\bibinfo{author}{\bibfnamefont{R.~E.} \bibnamefont{Goldstein}},
  \bibinfo{author}{\bibfnamefont{M.}~\bibnamefont{Polin}}, \bibnamefont{and}
  \bibinfo{author}{\bibfnamefont{I.}~\bibnamefont{Tuval}},
  \bibinfo{journal}{Phys. Rev. Lett.} \textbf{\bibinfo{volume}{103}},
  \bibinfo{pages}{168103} (\bibinfo{year}{2009}).

\bibitem[{\citenamefont{Woolley et~al.}(2009)\citenamefont{Woolley, Crockett,
  Groom, and Revell}}]{Woolley2009}
\bibinfo{author}{\bibfnamefont{D.~M.} \bibnamefont{Woolley}},
  \bibinfo{author}{\bibfnamefont{R.~F.} \bibnamefont{Crockett}},
  \bibinfo{author}{\bibfnamefont{W.~D.} \bibnamefont{Groom}}, \bibnamefont{and}
  \bibinfo{author}{\bibfnamefont{S.~G.} \bibnamefont{Revell}},
  \bibinfo{journal}{J. exp. Biol.} \textbf{\bibinfo{volume}{212}},
  \bibinfo{pages}{2215} (\bibinfo{year}{2009}).

\bibitem[{\citenamefont{Brumley et~al.}(2014)\citenamefont{Brumley, Wan, Polin,
  and Goldstein}}]{Brumley2014}
\bibinfo{author}{\bibfnamefont{D.~R.} \bibnamefont{Brumley}},
  \bibinfo{author}{\bibfnamefont{K.~Y.} \bibnamefont{Wan}},
  \bibinfo{author}{\bibfnamefont{M.}~\bibnamefont{Polin}}, \bibnamefont{and}
  \bibinfo{author}{\bibfnamefont{R.~E.} \bibnamefont{Goldstein}},
  \bibinfo{journal}{eLife} \textbf{\bibinfo{volume}{3}},
  \bibinfo{pages}{5030732} (\bibinfo{year}{2014}).

\bibitem[{\citenamefont{Pellicciotta et~al.}(2020)\citenamefont{Pellicciotta,
  Hamilton, Kotar, Faucourt, Delgehyr, Spassky, and Cicuta}}]{Pellicciotta2020}
\bibinfo{author}{\bibfnamefont{N.}~\bibnamefont{Pellicciotta}},
  \bibinfo{author}{\bibfnamefont{E.}~\bibnamefont{Hamilton}},
  \bibinfo{author}{\bibfnamefont{J.}~\bibnamefont{Kotar}},
  \bibinfo{author}{\bibfnamefont{M.}~\bibnamefont{Faucourt}},
  \bibinfo{author}{\bibfnamefont{N.}~\bibnamefont{Delgehyr}},
  \bibinfo{author}{\bibfnamefont{N.}~\bibnamefont{Spassky}}, \bibnamefont{and}
  \bibinfo{author}{\bibfnamefont{P.}~\bibnamefont{Cicuta}},
  \bibinfo{journal}{Proc. Natl. Acad. Sci. U.S.A.}
  \textbf{\bibinfo{volume}{117}}, \bibinfo{pages}{8315} (\bibinfo{year}{2020}).

\bibitem[{\citenamefont{Purchell}(1977)}]{Purcell1977}
\bibinfo{author}{\bibfnamefont{E.~M.} \bibnamefont{Purchell}},
  \bibinfo{journal}{Amer. J. Phys.} \textbf{\bibinfo{volume}{45}},
  \bibinfo{pages}{3} (\bibinfo{year}{1977}).

\bibitem[{\citenamefont{Lauga and Powers}(2009)}]{Lauga2009}
\bibinfo{author}{\bibfnamefont{E.}~\bibnamefont{Lauga}} \bibnamefont{and}
  \bibinfo{author}{\bibfnamefont{T.~R.} \bibnamefont{Powers}},
  \bibinfo{journal}{Rep. Progress Phys.} \textbf{\bibinfo{volume}{72}},
  \bibinfo{pages}{096601} (\bibinfo{year}{2009}).

\bibitem[{\citenamefont{Elgeti et~al.}(2015)\citenamefont{Elgeti, Winkler, and
  Gompper}}]{Elgeti2015}
\bibinfo{author}{\bibfnamefont{J.}~\bibnamefont{Elgeti}},
  \bibinfo{author}{\bibfnamefont{R.~G.} \bibnamefont{Winkler}},
  \bibnamefont{and} \bibinfo{author}{\bibfnamefont{G.}~\bibnamefont{Gompper}},
  \bibinfo{journal}{Rep. Progress. Phys.} \textbf{\bibinfo{volume}{78}},
  \bibinfo{pages}{056601} (\bibinfo{year}{2015}).

\bibitem[{\citenamefont{Happel and Brenner}(1965)}]{Happel:hydro}
\bibinfo{author}{\bibfnamefont{J.}~\bibnamefont{Happel}} \bibnamefont{and}
  \bibinfo{author}{\bibfnamefont{H.}~\bibnamefont{Brenner}},
  \emph{\bibinfo{title}{{Low Reynolds Number Hydrodynamics}}}
  (\bibinfo{publisher}{Kluwer}, \bibinfo{address}{Boston, MA},
  \bibinfo{year}{1965}).

\bibitem[{\citenamefont{Gray and Hancock}(1955)}]{Gray1955b}
\bibinfo{author}{\bibfnamefont{J.}~\bibnamefont{Gray}} \bibnamefont{and}
  \bibinfo{author}{\bibfnamefont{G.~T.} \bibnamefont{Hancock}},
  \bibinfo{journal}{J. exp. Biol.} \textbf{\bibinfo{volume}{32}},
  \bibinfo{pages}{802} (\bibinfo{year}{1955}).

\bibitem[{\citenamefont{Johnson and Brokaw}(1979)}]{Johnson1979}
\bibinfo{author}{\bibfnamefont{R.}~\bibnamefont{Johnson}} \bibnamefont{and}
  \bibinfo{author}{\bibfnamefont{C.}~\bibnamefont{Brokaw}},
  \bibinfo{journal}{Biophys. J.} \textbf{\bibinfo{volume}{25}},
  \bibinfo{pages}{113} (\bibinfo{year}{1979}).

\bibitem[{\citenamefont{Friedrich et~al.}(2010)\citenamefont{Friedrich,
  Riedel-Kruse, Howard, and J{\"{u}}licher}}]{Friedrich2010}
\bibinfo{author}{\bibfnamefont{B.~M.} \bibnamefont{Friedrich}},
  \bibinfo{author}{\bibfnamefont{I.~H.} \bibnamefont{Riedel-Kruse}},
  \bibinfo{author}{\bibfnamefont{J.}~\bibnamefont{Howard}}, \bibnamefont{and}
  \bibinfo{author}{\bibfnamefont{F.}~\bibnamefont{J{\"{u}}licher}},
  \bibinfo{journal}{J. exp. Biol.} \textbf{\bibinfo{volume}{213}},
  \bibinfo{pages}{1226} (\bibinfo{year}{2010}).

\bibitem[{\citenamefont{Batchelor}(1970)}]{Batchelor1970}
\bibinfo{author}{\bibfnamefont{G.}~\bibnamefont{Batchelor}},
  \bibinfo{journal}{J. Fluid Mech.} \textbf{\bibinfo{volume}{44}},
  \bibinfo{pages}{419} (\bibinfo{year}{1970}).

\bibitem[{\citenamefont{Keller and Rubinow}(1976)}]{Keller1976}
\bibinfo{author}{\bibfnamefont{J.~B.} \bibnamefont{Keller}} \bibnamefont{and}
  \bibinfo{author}{\bibfnamefont{S.~I.} \bibnamefont{Rubinow}},
  \bibinfo{journal}{J. Fluid Mech.} \textbf{\bibinfo{volume}{75}},
  \bibinfo{pages}{705} (\bibinfo{year}{1976}).

\bibitem[{\citenamefont{Smith}(2009)}]{Smith2009}
\bibinfo{author}{\bibfnamefont{D.~J.} \bibnamefont{Smith}},
  \bibinfo{journal}{Proc. Roy. Soc. A: Mathematical, Physical and Engineering
  Sciences} \textbf{\bibinfo{volume}{465}}, \bibinfo{pages}{3605}
  (\bibinfo{year}{2009}).

\bibitem[{\citenamefont{Gompper et~al.}(2009)\citenamefont{Gompper, Ihle,
  Kroll, and Winkler}}]{Gompper2009}
\bibinfo{author}{\bibfnamefont{G.}~\bibnamefont{Gompper}},
  \bibinfo{author}{\bibfnamefont{T.}~\bibnamefont{Ihle}},
  \bibinfo{author}{\bibfnamefont{D.}~\bibnamefont{Kroll}}, \bibnamefont{and}
  \bibinfo{author}{\bibfnamefont{R.}~\bibnamefont{Winkler}}, in
  \emph{\bibinfo{booktitle}{Advanced Computer Simulation Approaches for Soft
  Matter Sciences III}} (\bibinfo{publisher}{Springer}, \bibinfo{year}{2009}),
  pp. \bibinfo{pages}{1--87}.

\bibitem[{\citenamefont{Elgeti and Gompper}(2008)}]{Elgeti2008}
\bibinfo{author}{\bibfnamefont{J.}~\bibnamefont{Elgeti}} \bibnamefont{and}
  \bibinfo{author}{\bibfnamefont{G.}~\bibnamefont{Gompper}}, in
  \emph{\bibinfo{booktitle}{NIC Symposium}} (\bibinfo{year}{2008}),
  vol.~\bibinfo{volume}{39}, pp. \bibinfo{pages}{53--62}.

\bibitem[{\citenamefont{Winkler}(2016)}]{Winkler2016}
\bibinfo{author}{\bibfnamefont{R.~G.} \bibnamefont{Winkler}},
  \bibinfo{journal}{Europ. Phys. J. Spec. Topics}
  \textbf{\bibinfo{volume}{225}}, \bibinfo{pages}{2079} (\bibinfo{year}{2016}).

\bibitem[{\citenamefont{Westphal et~al.}(2014)\citenamefont{Westphal, Singh,
  Huang, Gompper, and Winkler}}]{Westphal2014}
\bibinfo{author}{\bibfnamefont{E.}~\bibnamefont{Westphal}},
  \bibinfo{author}{\bibfnamefont{S.~P.} \bibnamefont{Singh}},
  \bibinfo{author}{\bibfnamefont{C.-C.} \bibnamefont{Huang}},
  \bibinfo{author}{\bibfnamefont{G.}~\bibnamefont{Gompper}}, \bibnamefont{and}
  \bibinfo{author}{\bibfnamefont{R.~G.} \bibnamefont{Winkler}},
  \bibinfo{journal}{Computer Physics Communications}
  \textbf{\bibinfo{volume}{185}}, \bibinfo{pages}{495} (\bibinfo{year}{2014}).

\bibitem[{\citenamefont{Chen and Doolen}(1998)}]{Chen1998}
\bibinfo{author}{\bibfnamefont{S.}~\bibnamefont{Chen}} \bibnamefont{and}
  \bibinfo{author}{\bibfnamefont{G.~D.} \bibnamefont{Doolen}},
  \bibinfo{journal}{Ann. Rev. Fluid Mech.} \textbf{\bibinfo{volume}{30}},
  \bibinfo{pages}{329} (\bibinfo{year}{1998}).

\bibitem[{\citenamefont{Liu and Nishimura}(2006)}]{Liu2006}
\bibinfo{author}{\bibfnamefont{Y.}~\bibnamefont{Liu}} \bibnamefont{and}
  \bibinfo{author}{\bibfnamefont{N.}~\bibnamefont{Nishimura}},
  \bibinfo{journal}{Engineering Analysis with Boundary Elements}
  \textbf{\bibinfo{volume}{30}}, \bibinfo{pages}{371} (\bibinfo{year}{2006}).

\bibitem[{\citenamefont{Liu}(2009)}]{Liu2009}
\bibinfo{author}{\bibfnamefont{Y.}~\bibnamefont{Liu}},
  \emph{\bibinfo{title}{Fast Multipole Boundary Element Method: Theory and
  Applications in Engineering}} (\bibinfo{publisher}{Cambridge University
  Press}, \bibinfo{address}{Cambridge ; New York}, \bibinfo{year}{2009}).

\bibitem[{\citenamefont{Giuliani et~al.}(2020)\citenamefont{Giuliani, Hess,
  DeSimone, and Rozza}}]{Giuliani2020}
\bibinfo{author}{\bibfnamefont{N.}~\bibnamefont{Giuliani}},
  \bibinfo{author}{\bibfnamefont{M.~W.} \bibnamefont{Hess}},
  \bibinfo{author}{\bibfnamefont{A.}~\bibnamefont{DeSimone}}, \bibnamefont{and}
  \bibinfo{author}{\bibfnamefont{G.}~\bibnamefont{Rozza}},
  \bibinfo{journal}{arXiv preprint arXiv:2006.13836}  (\bibinfo{year}{2020}).

\bibitem[{\citenamefont{Goldstein et~al.}(2002)\citenamefont{Goldstein, Poole,
  and Safko}}]{Goldstein:mechanics}
\bibinfo{author}{\bibfnamefont{H.}~\bibnamefont{Goldstein}},
  \bibinfo{author}{\bibfnamefont{C.}~\bibnamefont{Poole}}, \bibnamefont{and}
  \bibinfo{author}{\bibfnamefont{J.}~\bibnamefont{Safko}},
  \emph{\bibinfo{title}{{Classical Mechanics}}} (\bibinfo{year}{2002}).

\bibitem[{\citenamefont{Vilfan and Stark}(2009)}]{Vilfan2009}
\bibinfo{author}{\bibfnamefont{A.}~\bibnamefont{Vilfan}} \bibnamefont{and}
  \bibinfo{author}{\bibfnamefont{H.}~\bibnamefont{Stark}},
  \bibinfo{journal}{Phys. Rev. Lett.} \textbf{\bibinfo{volume}{103}},
  \bibinfo{pages}{199801} (\bibinfo{year}{2009}).

\bibitem[{\citenamefont{Friedrich and J\"ulicher}(2012)}]{Friedrich2012}
\bibinfo{author}{\bibfnamefont{B.~M.} \bibnamefont{Friedrich}}
  \bibnamefont{and}
  \bibinfo{author}{\bibfnamefont{F.}~\bibnamefont{J\"ulicher}},
  \bibinfo{journal}{Phys. Rev. Lett.} \textbf{\bibinfo{volume}{109}},
  \bibinfo{pages}{138102} (\bibinfo{year}{2012}).

\bibitem[{\citenamefont{Geyer et~al.}(2013)\citenamefont{Geyer, J{\"u}licher,
  Howard, and Friedrich}}]{Geyer2013}
\bibinfo{author}{\bibfnamefont{V.~F.} \bibnamefont{Geyer}},
  \bibinfo{author}{\bibfnamefont{F.}~\bibnamefont{J{\"u}licher}},
  \bibinfo{author}{\bibfnamefont{J.}~\bibnamefont{Howard}}, \bibnamefont{and}
  \bibinfo{author}{\bibfnamefont{B.~M.} \bibnamefont{Friedrich}},
  \bibinfo{journal}{Proc. Natl. Acad. Sci. U.S.A.}
  \textbf{\bibinfo{volume}{110}}, \bibinfo{pages}{18058}
  (\bibinfo{year}{2013}).

\bibitem[{\citenamefont{Polotzek and Friedrich}(2013)}]{Polotzek2013}
\bibinfo{author}{\bibfnamefont{K.}~\bibnamefont{Polotzek}} \bibnamefont{and}
  \bibinfo{author}{\bibfnamefont{B.~M.} \bibnamefont{Friedrich}},
  \bibinfo{journal}{New J. Phys.} \textbf{\bibinfo{volume}{15}},
  \bibinfo{pages}{045005} (\bibinfo{year}{2013}).

\bibitem[{\citenamefont{Klindt and Friedrich}(2015)}]{Klindt2015}
\bibinfo{author}{\bibfnamefont{G.}~\bibnamefont{Klindt}} \bibnamefont{and}
  \bibinfo{author}{\bibfnamefont{B.}~\bibnamefont{Friedrich}},
  \bibinfo{journal}{Phys. Rev. E} \textbf{\bibinfo{volume}{92}}
  (\bibinfo{year}{2015}).

\bibitem[{\citenamefont{Klindt et~al.}(2017)\citenamefont{Klindt, Ruloff,
  Wagner, and Friedrich}}]{Klindt2017}
\bibinfo{author}{\bibfnamefont{G.~S.} \bibnamefont{Klindt}},
  \bibinfo{author}{\bibfnamefont{C.}~\bibnamefont{Ruloff}},
  \bibinfo{author}{\bibfnamefont{C.}~\bibnamefont{Wagner}}, \bibnamefont{and}
  \bibinfo{author}{\bibfnamefont{B.~M.} \bibnamefont{Friedrich}},
  \bibinfo{journal}{New J. Phys.} \textbf{\bibinfo{volume}{19}},
  \bibinfo{pages}{113052} (\bibinfo{year}{2017}).

\bibitem[{\citenamefont{Taylor}(1951)}]{Taylor1951}
\bibinfo{author}{\bibfnamefont{G.~I.} \bibnamefont{Taylor}},
  \bibinfo{journal}{Proc. Roy. Soc. A. Mathematical and Physical Sciences}
  \textbf{\bibinfo{volume}{209}}, \bibinfo{pages}{447} (\bibinfo{year}{1951}).

\bibitem[{\citenamefont{Najafi and Golestanian}(2004)}]{Najafi2004}
\bibinfo{author}{\bibfnamefont{A.}~\bibnamefont{Najafi}} \bibnamefont{and}
  \bibinfo{author}{\bibfnamefont{R.}~\bibnamefont{Golestanian}},
  \bibinfo{journal}{Phys. Rev. E} \textbf{\bibinfo{volume}{69}},
  \bibinfo{pages}{062901} (\bibinfo{year}{2004}).

\bibitem[{\citenamefont{Becker et~al.}(2003)\citenamefont{Becker, Koehler, and
  Stone}}]{Becker2003}
\bibinfo{author}{\bibfnamefont{L.~E.} \bibnamefont{Becker}},
  \bibinfo{author}{\bibfnamefont{S.~A.} \bibnamefont{Koehler}},
  \bibnamefont{and} \bibinfo{author}{\bibfnamefont{H.~A.} \bibnamefont{Stone}},
  \bibinfo{journal}{J. Fluid Mech.} \textbf{\bibinfo{volume}{490}},
  \bibinfo{pages}{15} (\bibinfo{year}{2003}).

\bibitem[{\citenamefont{Dreyfus et~al.}(2005)\citenamefont{Dreyfus, Baudry, and
  Stone}}]{Dreyfus2005}
\bibinfo{author}{\bibfnamefont{R.}~\bibnamefont{Dreyfus}},
  \bibinfo{author}{\bibfnamefont{J.}~\bibnamefont{Baudry}}, \bibnamefont{and}
  \bibinfo{author}{\bibfnamefont{H.~A.} \bibnamefont{Stone}},
  \bibinfo{journal}{Europ. Phys. J. B - Condensed Matter and Complex Systems}
  \textbf{\bibinfo{volume}{47}}, \bibinfo{pages}{161} (\bibinfo{year}{2005}).

\bibitem[{\citenamefont{Ma et~al.}(2014)\citenamefont{Ma, Klindt, Riedel-Kruse,
  J{\"u}licher, and Friedrich}}]{Ma2014}
\bibinfo{author}{\bibfnamefont{R.}~\bibnamefont{Ma}},
  \bibinfo{author}{\bibfnamefont{G.~S.} \bibnamefont{Klindt}},
  \bibinfo{author}{\bibfnamefont{I.~H.} \bibnamefont{Riedel-Kruse}},
  \bibinfo{author}{\bibfnamefont{F.}~\bibnamefont{J{\"u}licher}},
  \bibnamefont{and} \bibinfo{author}{\bibfnamefont{B.~M.}
  \bibnamefont{Friedrich}}, \bibinfo{journal}{Phys. Rev. Lett.}
  \textbf{\bibinfo{volume}{113}}, \bibinfo{pages}{048101}
  (\bibinfo{year}{2014}).

\bibitem[{\citenamefont{Wan and Goldstein}(2014)}]{Wan2014}
\bibinfo{author}{\bibfnamefont{K.~Y.} \bibnamefont{Wan}} \bibnamefont{and}
  \bibinfo{author}{\bibfnamefont{R.~E.} \bibnamefont{Goldstein}},
  \bibinfo{journal}{Phys. Rev. Lett.} \textbf{\bibinfo{volume}{113}},
  \bibinfo{pages}{238103} (\bibinfo{year}{2014}).

\bibitem[{\citenamefont{Werner et~al.}(2014)\citenamefont{Werner, Rink,
  Riedel-Kruse, and Friedrich}}]{Werner2014}
\bibinfo{author}{\bibfnamefont{S.}~\bibnamefont{Werner}},
  \bibinfo{author}{\bibfnamefont{J.~C.} \bibnamefont{Rink}},
  \bibinfo{author}{\bibfnamefont{I.~H.} \bibnamefont{Riedel-Kruse}},
  \bibnamefont{and} \bibinfo{author}{\bibfnamefont{B.~M.}
  \bibnamefont{Friedrich}}, \bibinfo{journal}{PLoS one}
  \textbf{\bibinfo{volume}{9}} (\bibinfo{year}{2014}).

\bibitem[{\citenamefont{Landau and Lifshitz}(1959)}]{Landau:fluid}
\bibinfo{author}{\bibfnamefont{L.~D.} \bibnamefont{Landau}} \bibnamefont{and}
  \bibinfo{author}{\bibfnamefont{E.~M.} \bibnamefont{Lifshitz}},
  \emph{\bibinfo{title}{Fluid Mechanics: Volume 6}}
  (\bibinfo{publisher}{Pergamon}, \bibinfo{year}{1959}).

\bibitem[{\citenamefont{Vilfan and J\"{u}licher}(2006)}]{Vilfan2006}
\bibinfo{author}{\bibfnamefont{A.}~\bibnamefont{Vilfan}} \bibnamefont{and}
  \bibinfo{author}{\bibfnamefont{F.}~\bibnamefont{J\"{u}licher}},
  \bibinfo{journal}{Phys. Rev. Lett.} \textbf{\bibinfo{volume}{96}},
  \bibinfo{pages}{58102} (\bibinfo{year}{2006}).

\bibitem[{\citenamefont{Niedermayer et~al.}(2008)\citenamefont{Niedermayer,
  Eckhardt, and Lenz}}]{Niedermayer2008}
\bibinfo{author}{\bibfnamefont{T.}~\bibnamefont{Niedermayer}},
  \bibinfo{author}{\bibfnamefont{B.}~\bibnamefont{Eckhardt}}, \bibnamefont{and}
  \bibinfo{author}{\bibfnamefont{P.}~\bibnamefont{Lenz}},
  \bibinfo{journal}{Chaos} \textbf{\bibinfo{volume}{18}},
  \bibinfo{pages}{037128} (\bibinfo{year}{2008}).

\bibitem[{\citenamefont{Uchida and Golestanian}(2011)}]{Uchida2011}
\bibinfo{author}{\bibfnamefont{N.}~\bibnamefont{Uchida}} \bibnamefont{and}
  \bibinfo{author}{\bibfnamefont{R.}~\bibnamefont{Golestanian}},
  \bibinfo{journal}{Phys, Rev. Lett.} \textbf{\bibinfo{volume}{106}},
  \bibinfo{pages}{058104} (\bibinfo{year}{2011}).

\bibitem[{\citenamefont{Murray et~al.}(1994)\citenamefont{Murray, Li, Sastry,
  and Sastry}}]{Murray:rigid}
\bibinfo{author}{\bibfnamefont{R.~M.} \bibnamefont{Murray}},
  \bibinfo{author}{\bibfnamefont{Z.}~\bibnamefont{Li}},
  \bibinfo{author}{\bibfnamefont{S.~S.} \bibnamefont{Sastry}},
  \bibnamefont{and} \bibinfo{author}{\bibfnamefont{S.~S.}
  \bibnamefont{Sastry}}, \emph{\bibinfo{title}{A mathematical introduction to
  robotic manipulation}} (\bibinfo{publisher}{CRC press},
  \bibinfo{year}{1994}).

\bibitem[{\citenamefont{Reichert}(2006)}]{Reichert:phd}
\bibinfo{author}{\bibfnamefont{M.}~\bibnamefont{Reichert}}, Ph.D. thesis,
  \bibinfo{school}{Universit{\"a}t Konstanz} (\bibinfo{year}{2006}).

\bibitem[{\citenamefont{Shapere and Wilczek}(1987)}]{Shapere1987}
\bibinfo{author}{\bibfnamefont{A.}~\bibnamefont{Shapere}} \bibnamefont{and}
  \bibinfo{author}{\bibfnamefont{F.}~\bibnamefont{Wilczek}},
  \bibinfo{journal}{Phys. Rev. Lett.} \textbf{\bibinfo{volume}{58}},
  \bibinfo{pages}{2051} (\bibinfo{year}{1987}).

\bibitem[{\citenamefont{Hinsen}(1995)}]{Hinsen1995}
\bibinfo{author}{\bibfnamefont{K.}~\bibnamefont{Hinsen}},
  \bibinfo{journal}{Computer Physics Communications}
  \textbf{\bibinfo{volume}{88}}, \bibinfo{pages}{327} (\bibinfo{year}{1995}).

\bibitem[{\citenamefont{Liu}(2020 (accessed September 28, 2020))}]{Liu:url}
\bibinfo{author}{\bibfnamefont{Y.}~\bibnamefont{Liu}},
  \emph{\bibinfo{title}{Yijun Liu's homepage}} (\bibinfo{year}{2020 (accessed
  September 28, 2020)}), \bibinfo{note}{http://www.yijunliu.com/}.

\bibitem[{\citenamefont{Blake}(1971)}]{Blake1971}
\bibinfo{author}{\bibfnamefont{J.~R.} \bibnamefont{Blake}},
  \bibinfo{journal}{Mathematical Proceedings of the Cambridge Philosophical
  Society} \textbf{\bibinfo{volume}{70}}, \bibinfo{pages}{303}
  (\bibinfo{year}{1971}).

\bibitem[{\citenamefont{Yan and Shelley}(2018)}]{Yan2018}
\bibinfo{author}{\bibfnamefont{W.}~\bibnamefont{Yan}} \bibnamefont{and}
  \bibinfo{author}{\bibfnamefont{M.}~\bibnamefont{Shelley}},
  \bibinfo{journal}{Journal of Computational Physics}
  \textbf{\bibinfo{volume}{375}}, \bibinfo{pages}{263} (\bibinfo{year}{2018}),
  \bibinfo{note}{zSCC: 0000001 arXiv: 1803.02424}.

\bibitem[{\citenamefont{Yan}(2020 (accessed September 28, 2020))}]{Yan:url}
\bibinfo{author}{\bibfnamefont{W.}~\bibnamefont{Yan}},
  \emph{\bibinfo{title}{STKFMM Github repository}} (\bibinfo{year}{2020
  (accessed September 28, 2020)}),
  \bibinfo{note}{https://github.com/wenyan4work/STKFMM}.

\bibitem[{\citenamefont{Naitoh and Sugino}(1984)}]{Naitoh1984}
\bibinfo{author}{\bibfnamefont{Y.}~\bibnamefont{Naitoh}} \bibnamefont{and}
  \bibinfo{author}{\bibfnamefont{K.}~\bibnamefont{Sugino}},
  \bibinfo{journal}{J. Protozoology} \textbf{\bibinfo{volume}{31}},
  \bibinfo{pages}{31} (\bibinfo{year}{1984}).

\bibitem[{\citenamefont{Guirao and Joanny}(2007)}]{Guirao2007}
\bibinfo{author}{\bibfnamefont{B.}~\bibnamefont{Guirao}} \bibnamefont{and}
  \bibinfo{author}{\bibfnamefont{J.-F.} \bibnamefont{Joanny}},
  \bibinfo{journal}{Biophys. J.} \textbf{\bibinfo{volume}{92}},
  \bibinfo{pages}{1900} (\bibinfo{year}{2007}).

\bibitem[{\citenamefont{Gueron and Levit-Gurevich}(1999)}]{Gueron1999}
\bibinfo{author}{\bibfnamefont{S.}~\bibnamefont{Gueron}} \bibnamefont{and}
  \bibinfo{author}{\bibfnamefont{K.}~\bibnamefont{Levit-Gurevich}},
  \bibinfo{journal}{Proc. Natl. Acad. Sci. U.S.A.}
  \textbf{\bibinfo{volume}{96}}, \bibinfo{pages}{12240} (\bibinfo{year}{1999}).

\bibitem[{\citenamefont{Elgeti and Gompper}(2013)}]{Elgeti2013}
\bibinfo{author}{\bibfnamefont{J.}~\bibnamefont{Elgeti}} \bibnamefont{and}
  \bibinfo{author}{\bibfnamefont{G.}~\bibnamefont{Gompper}},
  \bibinfo{journal}{Proc. Natl. Acad. Sci. U.S.A.}
  \textbf{\bibinfo{volume}{110}}, \bibinfo{pages}{4470} (\bibinfo{year}{2013}).

\bibitem[{\citenamefont{Wollin and Stark}(2011)}]{Wollin2011}
\bibinfo{author}{\bibfnamefont{C.}~\bibnamefont{Wollin}} \bibnamefont{and}
  \bibinfo{author}{\bibfnamefont{H.}~\bibnamefont{Stark}},
  \bibinfo{journal}{Europ. Phys. J. E} \textbf{\bibinfo{volume}{34}},
  \bibinfo{pages}{1} (\bibinfo{year}{2011}).

\bibitem[{\citenamefont{Kotar et~al.}(2013)\citenamefont{Kotar, Debono, Bruot,
  Box, Phillips, Simpson, Hanna, and Cicuta}}]{Kotar2013}
\bibinfo{author}{\bibfnamefont{J.}~\bibnamefont{Kotar}},
  \bibinfo{author}{\bibfnamefont{L.}~\bibnamefont{Debono}},
  \bibinfo{author}{\bibfnamefont{N.}~\bibnamefont{Bruot}},
  \bibinfo{author}{\bibfnamefont{S.}~\bibnamefont{Box}},
  \bibinfo{author}{\bibfnamefont{D.}~\bibnamefont{Phillips}},
  \bibinfo{author}{\bibfnamefont{S.}~\bibnamefont{Simpson}},
  \bibinfo{author}{\bibfnamefont{S.}~\bibnamefont{Hanna}}, \bibnamefont{and}
  \bibinfo{author}{\bibfnamefont{P.}~\bibnamefont{Cicuta}},
  \bibinfo{journal}{Phys. Rev. Lett.} \textbf{\bibinfo{volume}{111}},
  \bibinfo{pages}{228103} (\bibinfo{year}{2013}).

\bibitem[{\citenamefont{Meng et~al.}(2020)\citenamefont{Meng, Bennett, Uchida,
  and Golestanian}}]{Meng2020}
\bibinfo{author}{\bibfnamefont{F.}~\bibnamefont{Meng}},
  \bibinfo{author}{\bibfnamefont{R.~R.} \bibnamefont{Bennett}},
  \bibinfo{author}{\bibfnamefont{N.}~\bibnamefont{Uchida}}, \bibnamefont{and}
  \bibinfo{author}{\bibfnamefont{R.}~\bibnamefont{Golestanian}},
  \bibinfo{journal}{arXiv preprint arXiv:2007.02830}  (\bibinfo{year}{2020}).

\bibitem[{\citenamefont{Polin et~al.}(2009)\citenamefont{Polin, Tuval,
  Drescher, Gollub, and Goldstein}}]{Polin2009}
\bibinfo{author}{\bibfnamefont{M.}~\bibnamefont{Polin}},
  \bibinfo{author}{\bibfnamefont{I.}~\bibnamefont{Tuval}},
  \bibinfo{author}{\bibfnamefont{K.}~\bibnamefont{Drescher}},
  \bibinfo{author}{\bibfnamefont{J.~P.} \bibnamefont{Gollub}},
  \bibnamefont{and} \bibinfo{author}{\bibfnamefont{R.~E.}
  \bibnamefont{Goldstein}}, \bibinfo{journal}{Science}
  \textbf{\bibinfo{volume}{325}}, \bibinfo{pages}{487} (\bibinfo{year}{2009}).

\bibitem[{\citenamefont{Berg}(2008)}]{Berg:ecoli}
\bibinfo{author}{\bibfnamefont{H.~C.} \bibnamefont{Berg}},
  \emph{\bibinfo{title}{\textit{E. coli} in Motion}}
  (\bibinfo{publisher}{Springer Science \& Business Media},
  \bibinfo{year}{2008}).

\end{thebibliography}


\appendix

\renewcommand{\theequation}{S\arabic{equation}}    
\setcounter{equation}{0}  
\renewcommand{\thefigure}{S\arabic{figure}}    
\setcounter{figure}{0}  

\section{Appendix: Numerical methods}

We present additional details on the numerical computations for the application case of a pair of interacting cilia.

\begin{figure}
\includegraphics[width=0.7\linewidth]{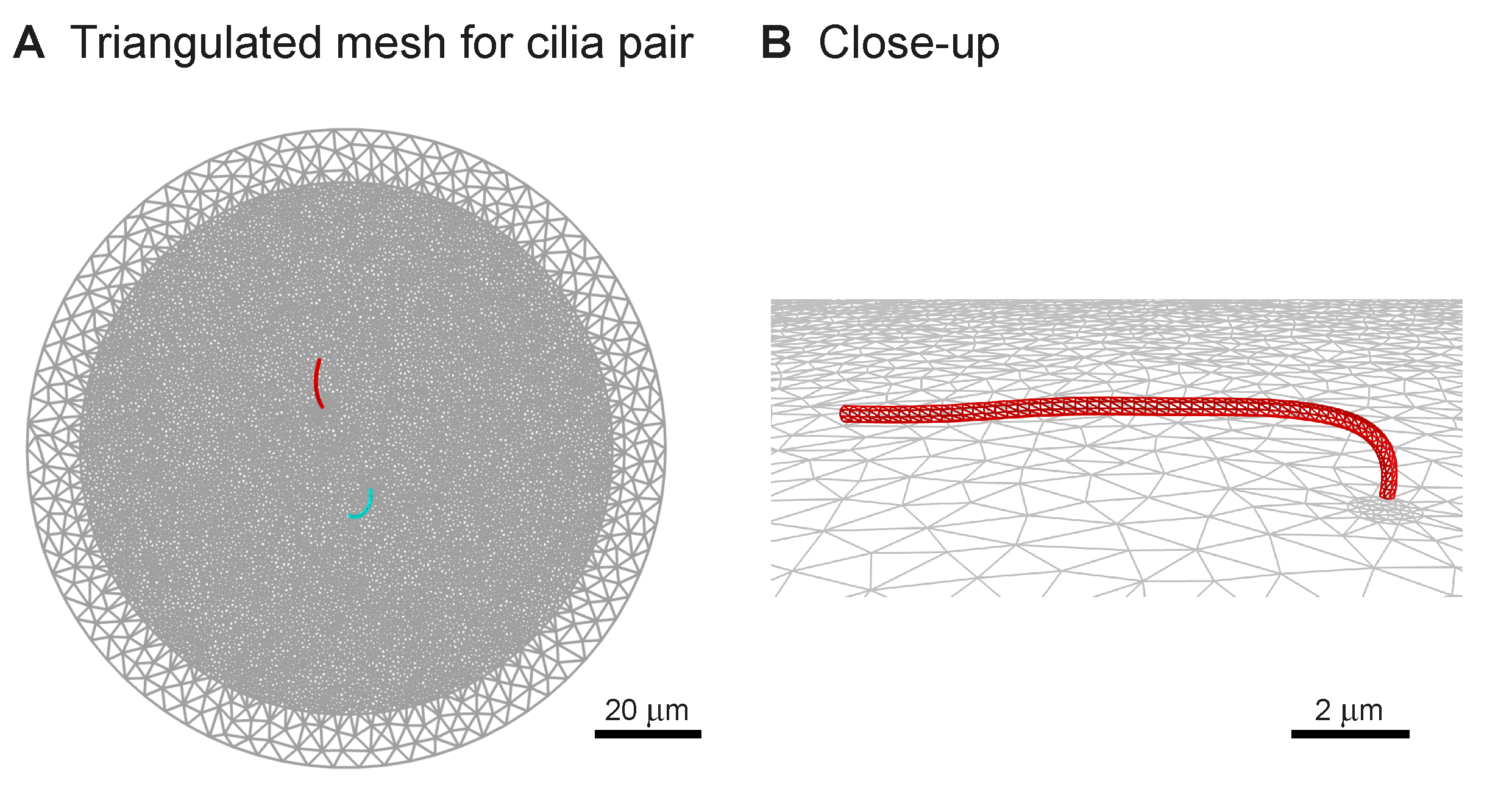}
\caption{
\textbf{Triangulated mesh for pair of cilia attached to boundary surface.}
(A)
Entire mesh consisting, of two cilia represented as bent cylinders (red, cyan),
as well as a thin disk of radius $60\,\micron$ representing the boundary surface (gray),
corresponding to approximately $1000$ triangular elements per cilium and $7500$ elements for the boundary surface.
(B)
Close-up view on a single cilium. 
Nodes on the bottom of the surface are hidden from view.
Cilia distance $d = 18\,\micron$, orientation angle $\psi = 2\pi/3$.
}
\label{figureS1}
\end{figure}

\paragraph{Mesh generation.}
We generated a triangulated mesh for the combined surface $\S$ of cilia and boundary surface 
using a custom-build Python package (available upon request), see also Fig.~\ref{figureS1}.
We represent the digitalized shapes of the cilia centerline
as a family of space curves $\r(s,\varphi)$ parameterized by arclength $s$ with $0\le s\le L$, 
where $L=10\,\micron$ is the length of the cilium, 
and a $2\pi$-periodic phase variable $\varphi$.
The centerline shapes of the two cilia are thus given by 
$\r_{0,1} + \r(s,\varphi_1)$ and
$\r_{0,2} + \r(s,\varphi_2)$, 
where the base points $\r_{0,1}$ and $\r_{0,2}$ 
have a distance $d$. 
The separation vector $\r_{0,2}-\r_{0,1}$ 
encloses an angle $\psi$ 
with $x$-axis (where $y$ axis is set by the effective stroke of both cilia), see Fig.~\ref{figure3}B. 

We generate a triangulated mesh for the each cilium by treating the cilium as a bent cylinder of radius $0.125\,\micron$, 
using 8 node points in azimuthal direction, and 61 nodes in longitudinal direction, 
as well as one apex node at the proximal and distal ends, respectively. 
For numerical stability, 
the proximal apices of each cilium mesh have a distance of $0.25\,\micron$ from the boundary surface.
A smaller distance virtually does not change the computed friction coefficients, but can cause convergence issues.

The hydrodynamic solver \textit{fastBEM} requires closed surfaces, 
which prompted us to use a circular disk of finite thickness (radius $60\,\micron$, thickness $1.5\,\micron$)
instead of a plane surface.
Initial simulations confirmed that using a larger disk radius virtually did not change results.
Disk faces were meshed using the Python \textit{triangle} package
(minimum triangle angle $20^\circ$, 
maximum triangle area $2\,\micron^2$ on the upper surface of the disk up to a distance of $50\,\micron$ from the disk center,
$20\,\micron^2$ otherwise).
Additionally, to improve the convergence of the solver, 
we refined the mesh in a small area below the proximal apices of the cilia
(maximum triangle area $0.04\,\micron^2$ up to a distance of $0.625\,\micron$ from cilia base points). 
In total,
each meshed cilium consists of 975 triangles, while
the meshed disk consists of approximately 7600 triangles.

\paragraph{Hydrodynamic computations.}

We employed a fast multipole boundary element method termed \textit{fastBEM} \cite{Liu2006,Liu2009, Liu:url} 
to solve the inverse
problem of finding the surface distributions of hydrodynamic friction forces  $\f(\x)$
for given given velocity fields $\v(\x)$ on the combined surface $\S$ of both cilia and the boundary surface.
To solve this inverse problem, 
the algorithm employs an iterative linear GMRES solver 
(tolerance parameter used here, $\mathrm{tol}=5 \cdot 10^{-4}$).
In principle, the solver would allow also for mixed boundary conditions 
that specify a combination of forces and velocities on different parts of the surface, which is, however, not needed here.

Initial tests showed that the self-friction is virtually independent of the phase the other cilium,
allowing us to approximate
$\Gamma_{11}(\varphi_1,\varphi_2) \approx \Gamma_{11}(\varphi_1)$ 
and
$\Gamma_{22}(\varphi_1,\varphi_2) \approx \Gamma_{11}(\varphi_2)$,
where $\Gamma_{11}(\varphi_1)$ corresponds to the simulation result for a single cilium.
For the smallest distance tested here, $14\,\micron$, the difference was at most $2\%$.
Thus, computation of $\Gamma_{11}$ required $m$ hydrodynamic computations for $m=20$ equidistant phase values.
The symmetry relation Eq.~\eqref{eq:gamma_symmetry} gives
$\Gamma_{21}(\varphi_1,\varphi_2) = \Gamma_{12}(\varphi_1,\varphi_2)$; 
thus it is sufficient to compute only $\Gamma_{12}$ 
(i.e., perform only computations where cilium number $2$ moves, while cilium number $1$ is static).
To compute $\Gamma_{12}(\varphi_1, \varphi_2)$,
we performed $m^2 = 400$ hydrodynamic computations
for $m^2$ pairs of phase values on a equidistant $(\varphi_1,\varphi_2)$-grid,
with mean CPU time of about $10^2$~seconds per computation.
We repeated these computations for $42$ different relative positions of cilia
as shown in Fig.~\ref{figure3}F.

\paragraph{Interpolation.}
From the generalized friction coefficients computed for a discrete set of $(\varphi_1,\varphi_2)$-values,
we obtained in a final step a continuous representation in the form of a (double) Fourier series truncated after order $4$
(corresponding, e.g., to $(2 \cdot 4+1)^2=81$ Fourier terms for $\Gamma_{12}(\varphi_1,\varphi_2)$).

\paragraph{Dynamical equation.}
The system of coupled ordinary differential equations, Eq.~\eqref{eq:cilia_pair},  
was solved with Python (method \textit{scipy.integrate.solve\_ivp}, tolerance $10^{-8}$, \textit{scipy} version 1.5.0).
In each time step, we compute the inverse of matrix $\mathbf{\Gamma}$.

\paragraph{Lyapunov exponents.}
For the computation of Lyapunov exponents  $\lambda$ shown in Fig.~\ref{figure3}F,
we used a small perturbation $\delta_0 = 10^{-2}$ of the in-phase synchronized state.
Preliminary simulations using a smaller perturbation $\delta_0 = 10^{-3}$ gave virtually identical results.

\paragraph{Steady-state phase difference.}
We computed the steady-state phase difference $\delta^\ast$ 
between the two cilia as a fixed point of the Poincar{\'e} map $\L:\delta_0\rightarrow\delta_1$.
Specifically, we computed $\L(\delta_0)$ for $30$ equidistant values of $\delta_0$ in the interval $[0,2\pi)$
by integrating Eq.~\eqref{eq:cilia_pair} using initial conditions $\varphi_1(t=0)=-\delta_0/2$ and $\varphi_2(t=0)=+\delta_0/2$.
We then numerically solved for fixed points $\L(\delta^\ast) = \delta^\ast$, 
using monotonic cubic spline interpolation of $\L$.
The periodic solution corresponding to a steady-state phase difference $\delta^\ast$ 
is stable if the numerical derivative 
$d \L/ d\delta_0|_{\delta_0=\delta^\ast}$
is smaller than $1$.

\begin{figure}
\includegraphics[width=0.99\linewidth]{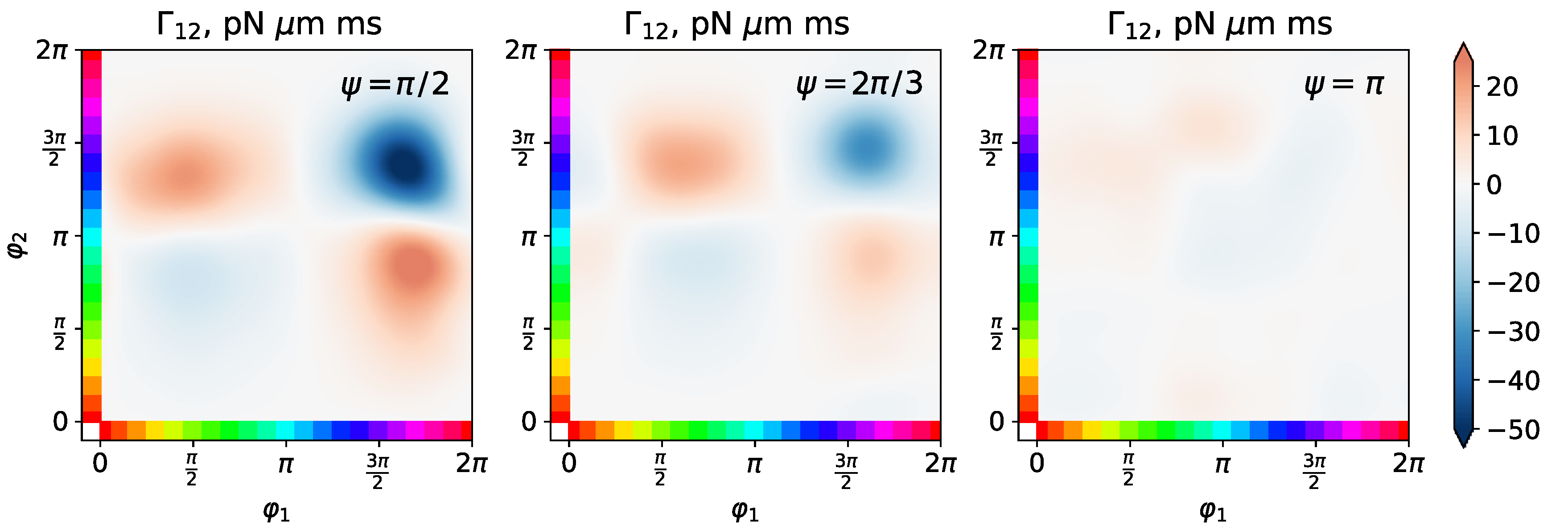}
\caption{
\textbf{Hydrodynamic interaction as function of cilia phases for different cilia orientations.}
Generalized hydrodynamic friction coefficient $\Gamma_{12}(\varphi_1, \varphi_2)$
as in Fig.~\ref{figure3}D for different cilia orientation angles:
left: $\psi = \pi/2$ (direction of effective stroke), 
middle: $\psi = 2\pi/3$ (oblique to direction of effective stroke),
right: $\psi = \pi$ (perpendicular to direction of effective stroke).
Cilia distance: $d = 18\,\micron$.
Note the different color scale compared to Fig.~\ref{figure3}D.
}
\label{figureS2}
\end{figure}

\end{document}